\documentclass[a4paper,11pt]{article}
\usepackage[utf8]{inputenc}
\usepackage[T1]{fontenc}
\usepackage[english]{babel}
\usepackage[title]{appendix}
\usepackage{authblk}
\usepackage{amsmath}
\usepackage{amssymb}
\usepackage{blindtext}
\usepackage{lipsum}
\usepackage{physics}
\usepackage{cite}
\usepackage{centernot}
\usepackage{wrapfig}
\usepackage{dsfont}
\usepackage{comment}
\usepackage{amsthm}
\usepackage{xfrac}
\usepackage{xcolor}
\usepackage{graphicx}
\usepackage{enumitem}
\usepackage[margin=1in,footskip=0.25in]{geometry}
\usepackage{soul}
\usepackage{mathtools}
\usepackage{hyperref}
\hypersetup{colorlinks=true, allcolors=blue}

\providecommand*{\unit}[1]{\,\ifmmode
	\mathrm{\,#1}\else\textup{#1}\fi}
\newcommand{\daga}[1]{{#1}^{\dagger}}

\newcommand{\normt}[1]{\norm{#1}_1}
\newcommand{\normd}[1]{\norm{#1}_\diamond}
\newcommand{\Md}[1]{M_{#1}(\mathbb{C})}
\newcommand{\bs}[1]{\boldsymbol{#1}}
\DeclareMathAlphabet{\mt}{U}{mt}{m}{n}

{% begin code
	\color{blue}%
	}%
	{\color{black}}% end code 

\newtheorem*{theorem*}{Theorem}

\newtheorem*{corollary*}{Corollary}

\newtheorem{proposition}{Proposition}

\newtheorem{remark}{Remark}

\begin{document}
	\title{\textbf{Superactivation of memory effects in a classical Markov  environment}}
	\author{Fabio Benatti}
	\author{Giovanni Nichele\footnote{giovanni.nichele@phd.units.it}}
	\affil{\small\textit{Dipartimento di Fisica, Università degli Studi di Trieste, I-34151 Trieste, Italy \textit{and} \\ \textit{Istituto Nazionale di Fisica Nucleare, Sezione di Trieste, I-34151 Trieste, Italy}}}
	\date{}% It is always \today, today,
			\maketitle

\begin{abstract}
	We investigate a phenomenon known as Superactivation of Backflow of Information (SBFI); namely, the fact that the tensor product of a non-Markovian dynamics with itself exhibits Backflow of Information (BFI) from environment to system even if the single dynamics does not. Such an effect is witnessed by the non-monotonic behaviour of the Helstrom norm and emerges in the open dynamics of two independent, but statistically coupled, parties. 
	We physically interpret SBFI  by means of the discrete-time non-Markovian dynamics of two open qubits collisionally coupled to an
	environment described by a classical Markov chain. In such a scenario, SBFI can be ascribed to the decrease of the qubit-qubit-environment correlations in favour of those of the two qubits, only.
	We further prove that the same mechanism at the roots of SBFI also holds in a suitable continuous-time limit. We also show that SBFI does not require entanglement to be witnessed, but only the quantumness of the Helstrom ensemble.
\end{abstract}

\section{Introduction}

Much study has recently been devoted to open quantum systems beyond the so-called Markovian regime~\cite{ChrusReview22}, when memory effects, for instance due to strong coupling to the environment,  cannot be neglected.
In contrast to the classical case, many non-equivalent concepts of quantum non-Markovianity have been  
put forward, forming an intricate hierarchy~\cite{LiLi}. 
In particular, for a one-parameter family $\{\Lambda_t\}_{t\ge0}$ of completely positive and trace preserving (CPTP) maps, non-Markovianity can be characterized according to two major approaches, involving either the Divisibility of the dynamics or the notion of Information Flow.
An open dynamics $\{\Lambda_t\}_{t\ge0}$ is divisible if, for all $t \ge s\ge0$, there exists an intertwiner map $\Lambda_{t,s}$ such that $\Lambda_t=\Lambda_{t,s}\Lambda_s$. Then, $\Lambda_t$ is said to be (C)P-divisible if all $\Lambda_{t,s}$ are (C)PTP maps. 
Markovianity has often been identified with CP-divisibility~\cite{RHPmeasure,Wolf2008Assessing}.

Instead, the so-called BLP approach~\cite{BLP} identifies Markovianity with the monotonic decrease in time of the distinguishability of generic states $\rho$ and $\sigma$, namely
$\partial_t \normt{\Lambda_t[\Delta_{\mu}(\rho,\sigma)]}\le 0$,
where $\normt{X}=\textrm{Tr}(\sqrt{X^\dagger X})$ denotes the trace-norm, while $\Delta_\mu(\rho,\sigma)=\mu \rho-(1-\mu)\sigma$ is the so-called Helstrom matrix, $\mu=1/2$ retrieving the trace distance $\normt{\rho-\sigma}/2$. A revival of state distinguishability, signalled by $\partial_t \normt{\Lambda_t[\Delta_{\mu}(\rho,\sigma)]}> 0$, for some $t> 0$, is then interpreted as Backflow of Information (BFI) from the environment to the system. Though a full-fledged microscopic characterization is still missing, BFI is usually associated with information stored in the form of system-environment correlations or with changes in the environmental state~\cite{MazzolaPaternostro,BLPWitnessInitialSystemenvironment2011,Megier2021entropic, HolevoSkew, BAV_2018}. 

It is known that P-divisible families $\{\Lambda_t\}_{t\geq 0}$ cannot support BFI since the maps $\Lambda_{t,s}$ are contractive~\cite{ChrusManiscalco,GTDWissmannBreuerAmato}. On the other hand, if the maps $\Lambda_t$ are invertible, P-divisible, but not CP-divisible, 
then the maps $\Lambda_t\otimes\Lambda_t$ cannot be P-divisible and thus show BFI at the level of a bipartite system even if the single system dynamics does not~\cite{BenattiChrusFil,BenattiNichele2023}. We call such a phenomenon  Superactivation of Backflow of Information (SBFI).

If one wants to access the actual flows of  information, if any, between system and environment, possibly at the roots of BFI and SBFI, sticking to the reduced dynamics of the system only is useless. Rather, a certain degree of control over the compound system-environment dynamics is needed.
As such, the microscopic physical mechanisms behind BFI are necessarily  heavily model-dependent and still debated. For instance, when a classical environment acts as a control on the quantum open system, while in~\cite{Megier_etal,LoFranco, Buscemi2024} the BFI seems not to be associated to an actual flow of information,
it is instead so in~\cite{BAV_2018}. Different points of view also appear regarding the discrimination of classical vs quantum effects behind the flows of information, in particular when trying to identify genuinely quantum memories~\cite{banacki2023information, BackerStrunz2024}.

Collisional models, though providing a discrete-time description, appear a suitable tool for assessing emerging memory effects from the system-environment dynamics~\cite{CMreview,CampbellCollisionModelsOpen2021a,StrunzCM2016,FilippovCM2017,FilippovCorrelated}. 
In the following, we therefore investigate the physics of SBFI in a discrete-time dissipative dynamics 
$\Lambda_n\otimes\Lambda_n$, $n\geq 0$, of two qubits that we obtain by means of a collisional model.
An algebraic approach typical of quantum Markov Chains~\cite{Kummerer2000scattering} will allow us to explain the emergence of SBFI in terms
	of the strength of the environment correlations and to describe it
by means of the time-behaviour of the mutual information between the open system and the classical Markov chain as its environment.
The results are as follows:
\begin{enumerate}
	\item
	for two qubits collisionally and unitarily interacting with a Markov chain environment, SBFI appears if successive chain sites are sufficiently correlated, in which case, the mutual information decreases in discrete-time without changes in the Markov chain state.
	This shows that part of the information shared by the two-qubit  and the collisional environment is released to the two-qubits.
	Such a result benefits from the algebraic approach to the collisional models developed in the following which provides a general context where to accommodate 
	correlated environments as for instance those treated in~\cite{FilippovCM2017,FilippovCorrelated, RybarFilippovZiman,BernardesCM}.
	\item
	A suitable continuous-time limit of the discrete-time qubit dynamics is obtained by means of  a non-unitary dynamical coupling between system and classical chain. 
	Also in this case the mutual information shows a non-monotonic behaviour confirming the interpretation of the SBFI as a loss of correlations between system 
	and  environment to the advantage of the open system.
	\item
	Finally, we show that the general non-classical resource needed for the emergence of SBFI is solely the 
	quantumness of the Helstrom ensemble, with no need of entanglement. 
\end{enumerate}

\section{Markov chain environment} 

As emphasized in the Introduction, we interpret the collisional scenario within an algebraic quantum spin chain approach.
We choose the environment $E$ to consist of an infinite spin chain, each site $k$ supporting 
a same $d\times d$ matrix algebra: ${\mathcal A}_E^{(k)}={\mathcal A}$. 
Local algebras ${\mathcal A}_{\bs{i}_{[-a,b]}}^{[-a,b]}=\bigotimes_{k=-a}^b {\mathcal A}_{i_k}^{(k)}$, supported by intervals $[-a,b]$ of integers $-a\leq j\leq b$, 
are generated by tensor products of the form $A^{[-a,b]}_{\bs{i}_{[-a,b]}}=\bigotimes_{j=-a}^{b}A^{(j)}_{i_j}$, where the upper index $(j)$ indicates the site at which the operator $A_{i_j}$ is located. These local operators  can be embedded within the infinite chain
as $\mathds{1}_E^{-a-1]}\otimes A_{\bs{i}_{[-a,b]}}^{[-a,b]}\otimes\mathds{1}^{[b+1}$, where 
$\mathds{1}_E^{-a-1]}=\otimes_{k=-\infty}^{-a-1}\mathds{1}^{(k)}_E$ and $\mathds{1}_E^{[b+1}=\otimes_{k=b+1}^{+\infty}\mathds{1}^{(k)}_E$.
In the following, for sake of simplicity, we will omit the infinite tensor products $\mathds{1}_E^{-a-1]}$ and $\mathds{1}_E^{[b+1}$. 
The collisional environment will then be described by the quasi-local ($C^*$) algebra $\mathcal{A}_E$ obtained by the so-called inductive limit of the local algebras~\cite{Bratteli}. 

Further, states over the chain are all positive, normalized linear expectations $\omega_E:\mathcal{A}_E\to\mathbb{C}$, 
$\omega_E(\mathds{1})=1$. When restricted to the local algebras, these expectations are represented by density matrices $\rho^{[-a,b]}\in\mathcal{A}_E^{[-a,b]}$ such that:
\begin{equation}
	\label{localstates}
	\omega_E\Big(\bigotimes_{k=-a}^bA^{(k)}_{i_k}\Big)={\rm Tr}\Big(\rho_E^{[-a,b]}\bigotimes_{k=-a}^bA^{(k)}_{i_k}\Big)\ ,
\end{equation}
for all $-a\le b$ and $A^{(k)}_{i_k}\in \mathcal{A}$ at site $k$.
Vice versa, a family of  density matrices $\rho^{[-a,b]}\in{\mathcal A}_E^{[a,b]}$, gives rise to a state $\omega_E$ over the chain if,
for all $-a\leq b$, ${\rm Tr}_b\rho^{[-a,b]}=\rho^{[-a,b-1]}$, where  ${\rm Tr}_k$ defines the partial trace over the $k$-th site. Environment correlations are present over the subset $[-a,b]$ whenever the density matrix $\rho^{[-a,b]}$
does not factorize.
Moreover, if ${\rm Tr}_{-a}\rho_E^{[-a,b]}=\rho_E^{[-a+1,b]}=\rho_E^{[-a,b-1]}$, then the state $\omega_E$ is invariant , $\omega_E\circ\Theta=\omega_E$, under the shift to the right, 
\begin{equation}
	\label{shift}
	\Theta\Big(A^{(-a)}_{i_{-a}}\otimes\cdots\otimes A^{(b)}_{i_b}\Big)=A^{(-a+1)}_{i_{-a}}\otimes\cdots\otimes A^{(b+1)}_{i_b}\ .
\end{equation}

\begin{figure}[t]
	\centering
	\includegraphics[width=.60\linewidth]{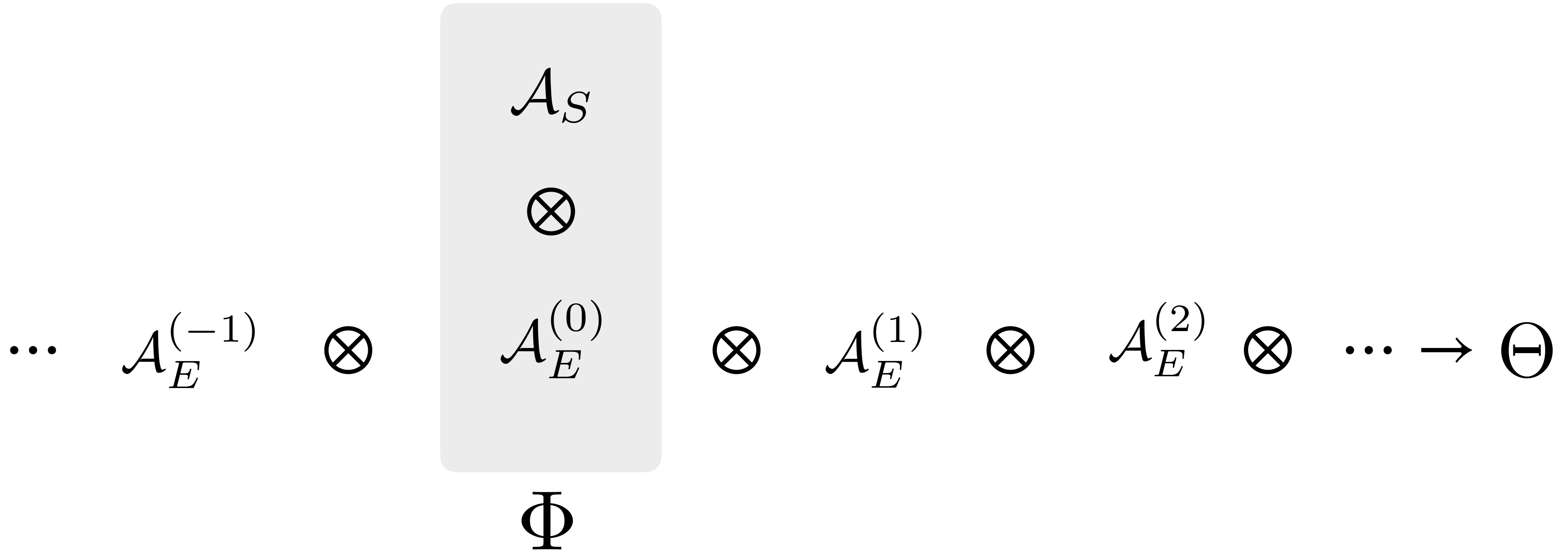}
	\caption{Scheme of the model for one party. The CPTP map $\Phi$ acts non-trivially only on the algebras of the system and of the $0$-th site of the chain, while $\Theta$ denotes the right shift on the chain. }
	\label{fig:scheme}
\end{figure}

We couple such a chain to a system $S$ described by a finite dimensional algebra $\mathcal{A}_S=\Md{\ell}$, with a state (expectation over 
$\mathcal{A}_S$) given by a density matrix $\rho_S$, $\omega_S(O_S)=\Tr(\rho_S\,O_S)$, $O_S\in\mathcal{A}_S$.  As depicted in 
Fig.~\ref{fig:scheme}, the $SE$ coupling is constructed as follows. 
Let $\Phi$ be a completely positive unital (CPU) map from $\mathcal{A_S}\otimes \mathcal{A}_E^{(0)}$ onto itself. Its action easily extends to the full algebra $\mathcal{A}_S\otimes \mathcal{A}_E$: 
\begin{eqnarray}\label{coupling}
	\Phi\Big[O_S\otimes A^{[-a,-1]}_{\bs{i}_{[-a,-1]}}\otimes A_{i_0}^{(0)}\otimes A_{\bs{i}_{[1,b]}}^{[1,b]}\Big] =A_{\bs{i}_{[-a,-1]}}^{[-a,-1]}\otimes
	\Phi[O_S\otimes A_{i_0}^{(0)}]\otimes A_{\bs{i}_{[1,b]}}^{[1,b]}\ .
\end{eqnarray}
The dynamics on the compound algebra $\mathcal{A}_S\otimes\mathcal{A}_E$ at discrete time $n$ is then given by  
\begin{equation}	
	\label{reduceddynamicsn}
	\Phi_n\equiv(\Theta\circ\Phi)^n \ .
\end{equation}
The maps $\Phi_n$ give the dynamics of operators  in the Heisenberg picture; in the Schr\"odinger picture, an initial state $\omega_{SE}$ on $\mathcal{A}_S\otimes\mathcal{A}_E$ evolves at discrete time $n$ into  
\begin{equation}
	\label{stteevolved}
	\omega_{SE}^{(n)}=\omega_{SE}\circ\Phi_n\ .
\end{equation}
Local restrictions to local algebras ${\mathcal A}_E^{[-a,b]}$ yield density matrices
$\Omega^{(n)}_{S[-a,b]}$ such that 
\begin{equation}
	\label{globstate}
	{\rm Tr}\Big(\Omega^{(n)}_{S[-a,b]}O_S\otimes A_E^{[-a,b]}\Big)
	=\omega_{SE}\circ\Phi_n\Big(O_S\otimes A^{[-a,b]}_E\Big)\ ,
\end{equation}
with marginal states
\begin{eqnarray}
	\label{marginal1}
	{\rm Tr}\Big(\Omega_S^{(n)}O_S\Big)&=& \omega_{SE}\circ\Phi_n\Big(O_S\otimes\mathds{1}_E\Big)\ ,\\  
	\label{marginal2}  
	{\rm Tr}\Big(\Omega^{(n)}_{[-a,b]}\,A_E^{[-a,b]}\Big)&=& \omega_{SE}\circ\Phi_n\Big(\mathds{1}_S\otimes A_E^{[-a,b]}\Big)\ ,
\end{eqnarray}
for all $O_S\in{\mathcal A}_S$ and $A_E^{[-a,b]}\in{\mathcal A}_E^{[-a,b]}$.

A factorized state $\omega_{SE}=\omega_S\otimes\omega_E$ on $\mathcal{A}_S\otimes\mathcal{A}_E$ is represented on 
$\mathcal{A}_S\otimes\mathcal{A}_E^{[-a,b]}$ by a factorized density matrix 
$\Omega_{S[-a,b]}=\rho_S\otimes\rho_E^{[-a,b]}$ and shows no correlations between system and collisional environment. Evidently, due to the dynamical coupling~\eqref{coupling}, correlations might  develop between $S$ and $E$ under the action of $\Phi_n$. Within the proposed algebraic setting, these correlations can be assessed by the mutual information.
For a generic bipartite system $A+B$ with state $\rho_{AB}$ and  marginals $\rho_{A,B}$ the mutual information is given by: 
\begin{equation}
	\label{mutualinf0}
	\mathcal{I}_{AB}=S(\rho_A)+S(\rho_B)-S(\rho_{AB})\geq 0\ ,
\end{equation}
where $S(\rho)=-{\rm Tr}\rho\log\rho$ is the von Neumann entropy.
Indeed, $S(\rho_{AB})\leq S(\rho_A)+S(\rho_B)$, while equality holds  if only if $\rho_{AB}=\rho_A\otimes\rho_B$.

As we are interested in the discrete-time behaviour of correlations between open system $S$ and sub-algebras $\mathcal{A}_E^{[-a,b]}$, we shall focus upon the following time-dependent
mutual information
\begin{equation}
	\label{mutualinf}
	\mathcal{I}^{(n)}_{S[-a,b]}=S\Big(\Omega^{(n)}_S\Big)+S\Big(\Omega^{(n)}_{[-a,b]}\Big)-
	S\Big(\Omega^{(n)}_{S[-a,b]}\Big)\ .
\end{equation}
An increase/decrease with $n$ of $\mathcal{I}^{(n)}_{S[-a,b]}$ would signal increasing/decreasing correlations between system and environment.

Let the environment $E$ be a commutative chain with at each site  a same commutative algebra $\mathcal{A}=D_d(\mathbb{C})$ spanned by $1$-dimensional orthogonal projections $\{\Pi_i\}_{i=0}^{d-1}$, $\sum_{k=0}^{d-1}\Pi_k=\mathds{1}$. It is turned into a Markov chain by endowing it with a state $\omega_E$, identified by the local density matrices: 
\begin{equation}
	\label{chainstate}
	\rho_E^{[-a,b]}=\sum_{\bs{i}_{[-a,b]}}p_{\bs{i}_{[-a,b]}}\,\Pi_{\bs{i}_{[-a,b]}}^{[-a,b]}\in\mathcal{A}_E^{[-a,b]}\ ,
\end{equation} 
where the projections $\Pi^{\otimes[-a,b]}_{\bs{i}_{[-a,b]}}=\bigotimes_{k=-a}^{b} \Pi_{i_k}^{(k)}$
generate the commutative sub-algebras $\mathcal{A}_E^{[-a,b]}$, and the probabilities $p_{\bs{i}_{[-a,b]}}$ satisfy:
\begin{equation}
	\label{Markovprop}
	p_{\bs{i}_{[-a,b]}}=T_{i_bi_{b-1}}\,T_{i_{b-1}i_{b-2}}\,\cdots\, T_{i_{-a+1}i_a}\,p_{i_{-a}}
\end{equation}
where $p_i\geq 0$, $\sum_{i=1}^dp_i=1$, while $T=[T_{ij}]$ satisfies $T_{ij}\geq 0$ and $\sum_{i=1}^dT_{ij}=1$ so that ${\rm Tr}_b\rho_E^{[a,b]}=\rho_E^{[a,b-1]}$.
Furthermore, the probability vector $\bs{p}=(p_1,\ldots,p_d)$ is chosen such that $T\bs{p}=\bs{p}$; then, ${\rm Tr}_a\rho^{[a,b]}_E=\rho_E^{[a+1,b]}=\rho_E^{[a,b-1]}$ and shift-invariance of the environment state is ensured, that is 
$\rho_E^{[a,b]}=\rho_E^{[a+n,b+n]}$ for all $n\in\mathbb{N}$.

Finally, let system and environment interact at site $0$ through the map 
\begin{equation}
	\label{collision}
	\Phi[O_S\otimes A_{i_0}^{(0)}]=\sum_{i=0}^{d-1}\phi_{i}[O_S] \otimes \Pi_i A_{i_0}^{(0)} \Pi_i\, ,
\end{equation}
the maps $\phi_{i}$ being completely positive and unital, $\phi_i[\mathds{1}]=\mathds{1}$. 
Then, as proved in Appendix~\ref{app:reduceddyn}, its extension to the whole tensor product ${\mathcal A}_S\otimes{\mathcal A}_E$ gives the step-$1$ dynamics
\begin{eqnarray}
	\nonumber
	\Phi_1[O_S\otimes A_{i_0}^{(0)}]&=&\Theta\circ\Phi[O_S\otimes A_{i_0}^{(0)}]\\
	\label{collisionshift}
	&=&\sum_{i=0}^{d-1}\phi_{i}[O_S] \otimes \Pi^{(1)}_i A^{(1)}_{i_0} \Pi^{(1)}_i\ .
\end{eqnarray}
%Notice that $\Phi[\mathds{1}_S\otimes A^{(0)}_{i_0}]= \mathds{1}_S\otimes A^{(0)}_{i_0}$; then,  the reduced state of the environment~\eqref{marginal2} is stationary; namely,
%\begin{equation}
%	\label{envstationarity}
%	\omega_{SE}(\Phi_n[\mathds{1}_S\otimes A^{[a,b]}_E])=\omega_{SE}(\mathds{1}_S\otimes A^{[a,b]}_E)\,.
%\end{equation} 
Furthermore, if the maps $\phi_i$ are invertible, 
$\Phi$ is an automorphism of the algebra $\mathcal{A}_S\otimes\mathcal{A}_E$; namely  $\Phi[AB]=\Phi[A]\Phi[B]$ for all $A,B\in\mathcal{A}_S\otimes\mathcal{A}_E$.

In summary, the algebraic setting just presented accommodates a collisional model within a correlated multi-partite classical environment~\cite{FilippovCM2017,FilippovCorrelated}, where system and ancilla at site $k=0$ may either interact reversibly 
or be instantaneously immersed in the same dissipative environment before the shift is applied.

Notice that $\Phi[\mathds{1}_S\otimes A^{(0)}_{i_0}]= \mathds{1}_S\otimes A^{(0)}_{i_0}$. As a consequence, the environment is stationary, 
\begin{equation}
	\label{envstationarity}
	\omega_{SE}(\Phi_n[\mathds{1}_S\otimes A^{[a,b]}_E])=\omega_{SE}(\mathds{1}_S\otimes A^{[a,b]}_E)\,.
\end{equation}  
Therefore, in the following, we focus upon the discrete-time reduced dynamics of the states of $S$, $\Lambda_n:\rho_S\mapsto\rho_{S_n}$. It is obtained from  restricting to the system $S$,
\begin{align}\label{partialexpectation_main}
	\omega_{SE} (\Phi_n[O_S\otimes \mathds{1}_E])=\Tr(\Lambda_n[\rho_S]O_S)\ .
\end{align}
We summarize the results concerning the reduced dynamics of system and environment in the following
\begin{proposition}\label{prop:dynmap}
	The reduced dynamics arising by collisional coupling~\eqref{collisionshift} of the system to a classical spin chain  in a state specified by~\eqref{chainstate} and~\eqref{Markovprop} 
	consists of a discrete-time family of CPTP maps,
	 \begin{equation}
	 	\hskip-.2cm\Lambda_n[\rho_S]:=
	 	\sum_{\bs{i}_{[1,n]}}p_{\bs{i}_{[1,n]}}^{\phantom{\ddagger}}\phi_{\bs{i}_{[1,n]}}^\ddagger[\rho_S]=\Omega_S^{(n)}\ ,\qquad \phi_{\bs{i}_{[1,n]}}^\ddagger=\phi_{i_n}^\ddagger\cdots\phi_{i_1}^\ddagger\ .
	 	\label{reduced_dyn} 
	 \end{equation}
	 with $\phi^\ddagger_i$ the CPTP map dual to the CPU map $\phi_i$ in~\eqref{collision}: ${\rm Tr}\Big(\rho_S\phi_i[O_S]\Big)={\rm Tr}\Big(\phi_i^\ddagger[\rho_s]O_S\Big)$.
	 On the other hand, the environment state is stationary,
	 \begin{equation}\label{stationary_envDM}
	 	\Omega_{[-a,b]}^{(n)}=\rho_E^{[-a,b]}\,.
	 \end{equation}
\end{proposition}
The proof is reported in Appendix~\ref{app:reduceddyn}. The collision model naturally provides a discrete-time dynamics. 
In discrete-time, the notion of divisibility is naturally drawn from continuous-time: $\Lambda_n$ is (C)P divisible if it can be written as  $\Lambda_n=\Lambda_{n,m}\, \Lambda_m$  $\forall \, n\ge m \in \mathbb{N},  $ with  $\Lambda_{n,m}=\Lambda_n^{\phantom{.}}\Lambda_m^{-1}$ a (C)PTP map. 

\subsection{Concrete collisional model} 

To investigate the physics behind the phenomenon of SBFI, we now consider two statistically coupled parties $S=S_1+S_2$, each independently interacting with its own Markov-chain environment, with compound reduced dynamics ${\Lambda_n\otimes\Lambda_n}$.
Thus $\mathcal{A}_S=M_2(\mathbb{C})\otimes M_2(\mathbb{C})$, while the Markov chain is chosen to consist of diagonal $4\times 4$ matrices, $\mathcal{A}_E^{(k)}=D_4(\mathbb{C})$, and 
$\phi_i$ to be unital Pauli maps:
\begin{equation}
	\label{Paulieig}   \phi_k[\sigma_j]=\mu_k^{(j)}\,\sigma_j\,,\qquad \mu_0^{(j)}=\mu^{(0)}_k=1\,,\qquad \mu^{(j)}_k=\varphi^{1-\delta_{jk}}\ ,
\end{equation}
for $j\neq 0,k\neq 0$ with $\varphi$ a real parameter, where  $\sigma_j$, $j=1,2,3$, are the Pauli matrices, while $\sigma_0=\mathds{1}$.
From~\eqref{reduced_dyn}, also $\Lambda_n$ results a unital Pauli map; indeed,
\begin{equation}
	\label{multeigen}
	\Lambda_n[\sigma_j]=\lambda_n^{(j)}\sigma_j\ ,\qquad \lambda_n^{(j)}=\sum_{\bs{i}_{[1,n]}}p_{\bs{i}_{[1,n]}}\,\mu^{(j)}_{\bs{i}_{[1,n]}}\ ,
\end{equation}
where $\mu^{(j)}_{\bs{i}_{[1,n]}}\equiv\prod_{k=1}^n\mu_{i_k}^{(j)}$.

The maps $\Lambda_n$ are invertible; then, 
$\Lambda_n=\Lambda_{n,n-1}\circ\Lambda_{n-1}$ with $\Lambda_{n,n-1}=\Lambda_n\circ\Lambda_{n-1}^{-1}$ and
\begin{equation}
	\label{intertw}
	\Lambda_{n,n-1}[\sigma_j]=\frac{\lambda_n^{(j)}}{\lambda^{(j)}_{n-1}}\,\sigma_j\ .
\end{equation}

Let the Markov transition $T$ in~\eqref{Markovprop} be
\begin{equation}\label{tmatrix}
	T=\begin{pmatrix}
		p_0 & p_0 & p_0 & p_0 \\
		p   & p+\Delta& p-\Delta & p\\
		p   & p-\Delta & p+\Delta & p\\ 
		r        & r& r     & r \\ 
	\end{pmatrix}	
\end{equation}
with positive parameters such that 
\begin{equation}
	\label{prob-const}
	0\le\Delta\le p\leq \frac{1}{2}\ ,\qquad p_0+2p+r=1\ ,
\end{equation}
and with invariant probability vector $\bs{p}=(p_0,p,p,r)$.
When $\Delta=0$, it follows that $T_{ij}=p_i$, for all $j$ so that the probabilities factorize, $p_{\bs{i}_{[-a,b]}}=\prod_{k=-a}^bp_{i_k}$, and  
$$
\rho_E^{[-a,b]}=\rho^{(-a)}_E\otimes\cdots\otimes\rho^{(b)}_E\ ,\qquad 
\rho^{(j)}_E=\sum_{i=0}^3 p_i\,\Pi^{(j)}_i\ .
$$
Further, from~\eqref{reduced_dyn}, it follows that such an uncorrelated environment  yields a reduced dynamics which is a CPTP discrete-time semigroup  $\Lambda_n^{\phantom{.}}=\Lambda^n$, where $\Lambda[\rho_S]=\sum_{i=0}^3 p_i\phi_i^\ddagger[\rho_S]$.

On the contrary, if $\Delta>0$, the mutual information in~\eqref{mutualinf0} with $\rho_A=\rho_E^{(k)}$, $\rho_B=\rho_E^{(k+1)}$ and $\rho_{AB}=\rho_E^{[k,k+1]}$ yields 
$$
\mathcal{I}_{k,k+1}=4p^2\left(\log 2-h\left(\frac{1+Q}{2}\right)\right)\ ,\qquad Q\equiv \frac{\Delta}{p}\,,
$$
and $h(x)=-x\log x-(1-x)\log(1-x)$ decreases for $1/2\leq x\leq 1$.  Due to the stationarity of the Markov process, $\mathcal{I}_{k,k+1}$ is site independent
and the correlations between any two successive environment sites increase with $0\leq\Delta\leq p$.
 Furthermore, for $\Delta>0$ the dynamical map $\Lambda_n$ is no longer a semigroup and the evolution is governed by the following
\begin{proposition}
	Choosing the maps $\phi_k$ as in \eqref{Paulieig} and the transition matrix as in~\eqref{tmatrix}, the spectrum of the dynamics $\Lambda_n$, 
	\begin{equation*}
		\Lambda_n[\sigma_j]=\lambda_n^{(j)}\sigma_j\ ,\qquad \lambda_n^{(j)}=\sum_{\bs{i}_{[1,n]}}p_{\bs{i}_{[1,n]}}\,\mu^{(j)}_{\bs{i}_{[1,n]}}\,,\qquad j=0,1,2,3\,,
	\end{equation*}
	satisfies the following recurrences
	\begin{eqnarray}\label{spectrumrecurrence}
		\lambda_n^{(1,2)}&=:&\lambda_n=[1-(p+r)(1-\varphi)]\lambda_{n-1}
		+p\,\Delta (1-\varphi)^2\,\sum_{j=0}^{n-2}\lambda_j\,[ (1+\varphi)\Delta]^{n-j-2}\, ,\\
		\label{spectrumrecurrence1}	
		\lambda_n^{(3)}&=&
		[1-2p\,(1-\varphi)]\,\lambda_{n-1}^{(3)}\ .
	\end{eqnarray}
\end{proposition}
\begin{proof}
	
	Due to the form of the transition matrix,
	\begin{equation*}\label{tmatrixapp}
		T=\begin{pmatrix}
			p_0 & p_0 & p_0 & p_0 \\
			p   & p& p& p\\
			p   & p & p & p\\ 
			r        & r& r     & r \\ 
		\end{pmatrix}\,+\,\Delta\,\begin{pmatrix}
			0 & 0 & 0 & 0 \\
			0   & 1& -1 & 0\\
			0   & -1 & 1 & 0\\ 
			0        & 0& 0     & 0 \\ 
		\end{pmatrix}	\ ,	
	\end{equation*}
	summing over the index $i_n$ in~\eqref{multeigen} yields
	\begin{equation}
		\label{appB1}
		\lambda^{(j)}_n=A^{(j)}_{n-1}\,\lambda^{(j)}_{n-1}+\Delta\Big(\mu^{(j)}_1-\mu^{(j)}_2\Big)\,B^{(j)}_{n-1}\qquad\forall j=0,1,2,3\ ,
	\end{equation}
	where, for $n\geq 1$
	\begin{equation}
		\label{appB2}
		A^{(j)}_{n-1}=p_0\,+\,p\,\Big(\mu^{(j)}_1+\mu^{(j)}_2\Big)\,+\,r\mu^{(j)}_3\, ,\quad
		B^{(j)}_{n-1}=\sum_{\bs{i}_{[1,n-2]}}\Big(T_{1i_{n-2}}\mu^{(j)}_1-T_{2i_{n-2}}\mu^{(j)}_2\Big)p_{\bs{i}_{[1,n-2]}}\,\mu^{(j)}_{\bs{i}_{[1,n-2]}}\, ,
	\end{equation}
	with $p_{\bs{i}^0}=1$, $T_{i0}=p_i$ and $B^{(j)}_0=0$.
	Then, summing over $i_{n-2}$ in the expression for $B^{(j)}_{n-1}$, one gets
	\begin{equation}
		\label{appB3}
		B^{(j)}_{n-1}=p\,\Big(\mu^{(j)}_1-\mu^{(j)}_2\Big)\,\lambda^{(j)}_{n-2}\,+\,\Delta\,\Big(\mu^{(j)}_1+\mu^{(j)}_2\Big)\,B^{(j)}_{n-2}\ ,
	\end{equation}
	and, iterating,
	\begin{eqnarray}
		\nonumber
		B^{(j)}_{n-1}
		&=&p\,\Big(\mu^{(j)}_1-\mu^{(j)}_2\Big)\,\Bigg(\lambda^{(j)}_{n-2}\,+\,\Delta\,\Big(\mu^{(j)}_1+\mu^{(j)}_2\Big)\,\lambda^{(j)}_{n-3}\Bigg)
		\,+\,\Delta^2\,\Big(\mu^{(j)}_1+\mu^{(j)}_2\Big)^2\,B^{(j)}_{n-3}\\
		&=&p\,\Big(\mu^{(j)}_1-\mu^{(j)}_2\Big)\,\sum_{k=0}^{n-2}\lambda^{(j)}_k\,\Delta^{n-k-2}\,\Big(\mu^{(j)}_1+\mu^{(j)}_2\Big)^{n-k-2}\ ,
		\label{appB4}
	\end{eqnarray}
	where we set $\lambda^{(j)}_0=1$. 
	Since $\mu^{(0)}_j=1$ for $j=0,1,2,3$, from~\eqref{prob-const} it follows that $A^{(j)}_{n-1}=p_0+2p+r=1$ and $B^{(j)}_{n-1}=0$ so that $\lambda^{(0)}_n=1$ for all $n\in\mathbb{N}$.
	On the other hand, the choice of the other coefficients
	$\mu^{(j)}_k$ in~\eqref{Paulieig} gives
	\begin{eqnarray}
		\label{aux2a}
		A^{(1,2)}_{n-1}&=&p_0+p(1+\varphi)+r\varphi\ ,\qquad B_{n-1}^{(1,2)}=\pm p(1-\varphi)\sum_{k=0}^{n-2}\lambda^{(1,2)}_k\Delta^{n-k-2}(1+\varphi)^{n-k-2},\\
		\label{aux2b}
		A^{(3)}_{n-1}&=&p_0+2p\varphi+r\ ,\hskip 1.7cm B^{(3)}_{n-1}=0\ .
	\end{eqnarray}
	Since $p_0+2p+r=1$ the expressions in~\eqref{spectrumrecurrence} and~\eqref{spectrumrecurrence1} follow.
\end{proof}

We shall now study the model for two distinct choices of $\varphi$ in~\eqref{Paulieig},  corresponding respectively to (1) a unitary coupling, discussed in Section~\ref{subsub:unitary}, for which the solution of~\eqref{spectrumrecurrence} can be analytically computed and  (2) a dissipative coupling, presented in Section~\ref{subsub:diss}, for which  the natural stroboscopic limit of collisional models~\cite{CMreview,FilippovCM2017,Altamirano2017} is analytically available and allows one to compare the continuous-time scenario with the discrete-time one. 
\subsubsection{Unitary case}\label{subsub:unitary}
Set $\varphi=-1$; then, $\phi_k[X]=\sigma_kX\sigma_k$ and 
the map~\eqref{collision} becomes a ``controlled-unitary'' typical of collisional models~\cite{RybarFilippovZiman,BernardesCM}. 
In this scenario, the interaction between system $S$ and the environment $E$ is described by means of a unitary matrix $U_\tau=e^{-i g\tau \sum_k \sigma_k \otimes\Pi_k}$ for a duration $\tau={\pi}/{2g}$: $\Phi[X]=U^\dag_{{\pi}/{2g}}\,X\,U_{\pi/{2g}}$.
Only $j=n-2$ contributes to the sum in~\eqref{spectrumrecurrence} 
and in Appendix~\ref{app:unitary}, the recurrence relations~\eqref{spectrumrecurrence} and~\eqref{spectrumrecurrence1} are shown to yield
\begin{align}\label{lambda_n}
	\lambda_n&=\left(\frac{\beta+\alpha}{2\beta}\right)\left(\frac{\beta+\alpha}{2}\right)^{n}+\left(\frac{\beta-\alpha}{2\beta}\right)\left(\frac{\alpha-\beta}{2}\right)^{n}  , \nonumber\\
	\lambda_{n}^{(3)}&=\,(1-4p)^n \ ,
\end{align}
where we set
\begin{equation}
	\label{params}
	\alpha\equiv1-2(p+r)\ ,\qquad \beta\equiv \sqrt{\alpha^2+16\,p\,\Delta}\ .
\end{equation}
Let $\alpha>0$ so that $\lambda_n^{(1,2)}>0$.
The type of divisibility of the reduced dynamics depends on the environment correlations as follows.
\begin{proposition}\label{prop:divisibilitydegree}
	\begin{enumerate}[noitemsep,topsep=0pt,parsep=0pt,partopsep=0pt]
		\item[(i)] $\Lambda_n$ is P-divisible if and only if 
		\begin{equation}\label{Pdiv}
			2p\Delta \le r \alpha+p\alpha\,,
		\end{equation}
		\item[(ii)] $\Lambda_n$ is CP-divisible if and only if
		\begin{equation}\label{CPdiv}
			2p\Delta\le r \alpha\,,
		\end{equation}
		\item[(iii)] $\Lambda_n\otimes\Lambda_n$ is P-divisible if and only if 
		\begin{equation}
			\label{2timesdiv}
			2p\Delta\le \alpha(r+p)-\frac{\alpha}{2}\Big(1-\sqrt{1-4p(1-2p)}\Big)\ .
		\end{equation}
	\end{enumerate}
\end{proposition}
For the proof, see Appendix~\ref{app:unitary}. 
Notice the strength of the environmental correlations $\Delta$ governs the divisibility degree of the reduced dynamics, in that (ii)$\Rightarrow$ (iii) $\Rightarrow$ (i); on the other hand,  (iii) $\not\Rightarrow$ (ii) (see Remark~\ref{rem:CPP} below).

To illustrate  how the intensity of the environmental correlations relates to the emergence of SBFI, consider  $r=0$ 
so that $2\Delta\leq\alpha$ and, by (i), $\Lambda_n$ is guaranteed to be P-divisible. Then, the discrete-time intertwiners $\Lambda_{n,m}$ are contractive
and forbids BFI for a single qubit.

Then, we consider $p\ll1$ and proceed with a perturbative analysis.
Given any  $X=\daga{X}\in\Md{2}$, one has that (see Appendix~\ref{app:unitary} for details) up to second order in $p$,  
$$
\normt{\Lambda_{n,n-1}[X]}-\normt{X}=-  K_1\,p+K_2(\Delta)\,p^2 + o(p^2) \,, 
$$
with $K_1\ge0$ and $K_2(\Delta)>0$ and no discrete-time dependence. 
Therefore, possible environment correlations ($\Delta\neq 0$) contribute with a positive second order term in the small parameter $p$; this latter cannot counteract the negative, correlation independent first order term which then makes the maps $\Lambda_{n,n-1}$ contractive for all time-steps $n$
in the regime $0\leq \Delta\leq p\ll1$, thus concretely showing why there cannot be BFI for one qubit: the single qubit state distinguishability can never increase in time.

On the other hand, considering now two qubits, again setting $r=0$, at leading order in $0\leq p\ll1$, the positivity condition~\eqref{2timesdiv} implies $\Delta/p\equiv Q\le 1/2$. 
Therefore, if $Q>1/2$, $ \Lambda_{n,n-1}\otimes \Lambda_{n,n-1}$ cannot be positive and is thus not contractive. Moreover, being $\Lambda_n\otimes\Lambda_n$ invertible, the collisional dynamics of two qubits certainly exhibits SBFI, namely increasing distinguishability as witnessed by a suitably constructed two-qubit Helstrom statistical ensemble through the corresponding Helstrom matrix. 
Also, the lack of positivity of $ \Lambda_{n,n-1}\otimes \Lambda_{n,n-1}$ for $Q>1/2$ is easily seen by acting on totally symmetric projector $P_2^+$. Indeed, as shown in Appendix~\ref{app:unitary}, 
\begin{equation*}
	\normt{\Lambda_{n,n-1}\otimes \Lambda_{n,n-1}[P_2^+]}-\normt{P_2^+}= 4p^2 \ (2\,Q-1 )>0\,,
\end{equation*}
hence $\Lambda_{n,n-1}\otimes\Lambda_{n,n-1}$ is non-contractive, hence not Positive.    
\vskip .5cm
\begin{remark}
	\label{rem:CPP}
	Unlike $\Lambda_t\otimes\Lambda_t$ in continuous time,  in discrete time $\Lambda_n\otimes\Lambda_n$ can be P-divisible even if \ $\Lambda_n$ is not CP-divisible. Indeed, the main result of~\cite{BenattiChrusFil} is  based on the existence of time-local generators. Thus, even if $\Lambda_n$ is not CP-divisible, $\Lambda_n\otimes\Lambda_n$ need not automatically display SBFI. 
	However, as we saw above, in our case SBFI  is triggered by sufficiently strong environment correlations that help to violate the inequality~\eqref{2timesdiv}.
\end{remark}
\vskip .5cm
We now study the single and two qubit information flows from and into the collisional environment  by means of the system-environment correlations as quantified by the mutual information. For that  we restrict  the system-environment state at discrete-time $n$, $\omega_{SE}^{(n)}$, on a local observable $O_{S}\otimes A_E^{[-a+1,b]}$, $a,b \in \mathbb{N}$. One thus retrieves the evolved local system-environment density matrix $\Omega_{S[-a+1,b]}^{(n)}$ through~\eqref{globstate}, given by (see Appendix~\ref{app:localdm} eq.~\eqref{OmegaSE})
\begin{equation}\label{localSE}
	\Omega_{S [-a,b]}^{(n)}=\sum_{\bs{\ell}_{[-n+1,b]}}  p_{\bs{\ell}_{[-a,b]}}\phi^\ddagger_{\bs{\ell}_{[-n+1,0]}}[\rho_S]\otimes 
	\Pi_{\bs{\ell}_{[-n+1,b]}}^{[-a,b]}\,. 
\end{equation}
We shall then consider the mutual information~\eqref{mutualinf}
relative to the compound state at discrete-time $n$~\eqref{localSE} as a faithful quantifier of the system-chain correlations. In Appendix~\ref{app:localdm}, Eq.~\eqref{mutualI}, it is shown that the latter quantity takes the form
\begin{align}\nonumber
	\mathcal{I}^{(n)}_{S[-a,b]} =S(\Lambda_n[\rho_S])-\sum_{ \bs{i}_{[1,n]}} p_{\bs{i}_{[1,n]}}  \ S\left(\phi^\ddagger_{\bs{i}_{[1,n]}}[\rho_S]\right)\,.
\end{align}
Note that the previous expression depends only on $n$ and not on the size of the portion of the chain considered.
Taking into account, as above, two independent qubits coupled to identical chains, the maximal mutual information of their local density matrix reads 
\begin{eqnarray}
\label{qmi_discrete}
\mathcal{I}_{(S+S)E}^{(n)}=	S(\Lambda_n\otimes\Lambda_n[\rho_{S+S}])
	\ - \ \sum_{\mathclap{\bs{i}_{[1,n]},
			\bs{k}_{[1,n]}}} p_{\bs{i}_{[1,n]}} p_{\bs{k}_{[1,n]}}  S\left(\phi^\ddagger_{\bs{i}_{[1,n]}}\otimes\phi^\ddagger_{\bs{k}_{[1,n]}}[\rho_{S+S}]\right)\ .
	\label{qmi_discrete2}
\end{eqnarray} 
In the case under consideration, the unital maps $\phi_i$ are unitary; thus~\eqref{qmi_discrete} yields
\begin{equation}
	\label{entr-unitary}
	\mathcal{I}_{(S+S)E}^{(n)}=S(\Lambda_n\otimes\Lambda_n[\rho_{S+S}])-S(\rho_{S+S})\ ;
\end{equation}
in particular, the variation of the mutual information between two discrete-times 
$n\ge m$ reduces to checking the behaviour of two-qubit entropy:
\begin{equation}
	\label{mutualfinitedifference}
	\Delta \mathcal{I}_{(S+S)E}^{(n,m)}
	\equiv S(\Lambda_n\otimes \Lambda_n[\rho_{S+S}])-S(\Lambda_m\otimes\Lambda_m[\rho_{S+S}])\ .
\end{equation}
Let us recall that the von Neumann entropy increases under PTP unital maps~\cite{AlbertiUhlmann1982,AnielloChrusc2016,MuellerHermes}; thus, when the unital single-qubit reduced dynamics is P-divisible, $\Delta \mathcal{I}_{SE}^{(n,m)}\ge0$. On the other hand, moving to two qubits, 
choose as a concrete instance
\begin{equation}
	\label{choicepar}
	r=0\ ,\quad \frac{1}{4}\le p\le\frac{1}{2}\ ,\quad \Delta=\frac{1-2p}{2}\le p
	\le\frac{1}{2}\ ,
\end{equation}
so that $\Lambda_n$ is P-divisible with \eqref{Pdiv} being saturated and the Pauli eigenvalues~\eqref{lambda_n} at the first two successive discrete-time steps satisfy $\lambda_1=\lambda_2=\alpha=1-2p$.
Further, choosing $p= 1/4+\epsilon$, $\epsilon\ll1$, one can perform a perturbative study and show that the two-qubit completely symmetric projector 
$\rho_{S+S}=P_2^+$ witnesses a decrease of the two-qubit von Neumann entropy (details can be found in Appendix~\ref{app:Xx}),
\begin{equation}\label{computationdeltaQMI}
	\Delta\mathcal{I}_{(S+S)E}^{(2,1)}= -4 \log({4}/{3}) \, \epsilon^2<0\ ,
\end{equation}
hence a decrease of system
environment correlations between the first and the second collision.

\subsubsection{Dissipative case and stroboscopic limit}\label{subsub:diss}
Let us now take $\varphi=e^{-2\gamma \tau}$, $\gamma,\tau>0$ so that $\phi_0=\mathrm{id}$ and for $k\ne0$ 
\begin{equation}
	\label{paulidiss}
	\phi_k=e^{ \tau\mathcal{L}_k}\,, \qquad \mathcal{L}_k[X]=\gamma\,(\sigma_kX\sigma_k-X)\ .
\end{equation}
In such case, our model is analogous to a collisional model in which the qubit $\mathcal{A}_S$ and the ancilla $\mathcal{A}_E^{(0)}$ undergo a joint  dissipative evolution $O_S\otimes O_E^{(0)}\mapsto e^{ \tau\mathbb{L}}[O_S\otimes O_E^{(0)}]$ for a time $\tau$, before the shift on the chain is applied 
(the form of the the GKLS generator $\mathbb{L}$ is reported in Appendix~\ref{app:strobo}). The Markov chain correlations contribute with memory effects on top of this Markovian semigroup dynamics and, moreover, it allows one to retrieve a continuous-time dissipative dynamics and compare BFI and SBFI within such a continuous context. The technique employed is the so-called stroboscopic limit defined by $\tau\to0$, $n\to\infty$, $n\tau \to t$.
Choosing $\Delta={e^{-\kappa \tau}}/{2}$, $p\to1/2$ and, straightforwardly,  $\lambda_t^{(3)}=e^{-2\gamma t}$, while the other two Pauli eigenvalues are both equal to the solution $\lambda_t$ of the integro-differential equation
\begin{equation}
	\dot\lambda_t =-\gamma\, \lambda_t + \gamma^2\, \int_0^t \dd{s} e^{-(\kappa+\gamma)(t-s)}\, \lambda_s\,,
\end{equation}
which yields (see Appendix~\ref{app:strobo}) 
\begin{align}
	\lambda_t=\	e^{-(\gamma +\frac{\kappa}{2}) t } \bigg[\cosh \left(K t\right) +\frac{\kappa}{2K} \sinh \left(Kt\right)\bigg]\,,
\end{align}
where $K\equiv  \sqrt{\kappa^2+4 \gamma ^2}/2$. We thus obtain a family of P-divisible Pauli dynamical maps, with generator ${\mathcal{L}_t[\rho]=\frac{1}{2}\sum_{i=1}^3 \gamma_t^{(i)}(\sigma_i \rho \sigma_i -\rho)}$ and rates
\begin{align}
	\gamma_t^{(1)}&= \gamma_t^{(2)}=\gamma \,, \\
	\gamma_t^{(3)}&=-\frac{2 \gamma ^2}{\sqrt{\kappa^2+4 \gamma ^2} \coth \left(\frac{1}{2} t \sqrt{\kappa^2+4 \gamma ^2}\right)+\kappa}\ ,
\end{align}
with $\gamma_t^{(3)}$ being negative at all times.

\paragraph{System-environment correlations}
Let us consider the case $\Delta=1/2$ and $p=1/2$. Notice that such case corresponds to $\kappa=0$ and  $\gamma_t^{(3)}=-\gamma\tanh(\gamma t)$, namely to the well known ``eternally'' non-Markovian evolution firstly discussed in~\cite{HallCanonical}. 

In such case, only two sequences $\bs{i}_{[1,n]}$ have non-vanishing probabilities and thus contribute to~\eqref{reduced_dyn}, namely $\bs{1}=1 1 1 \dots $ and $\bs{2}=2 2 2 \dots $ with probabilities $p_{\bs{1}}=p_{\bs{2}}={1}/{2}$. Accordingly, the continuous-time limit of~\eqref{qmi_discrete} reads
\begin{equation}\label{qmi_continuous}
	\mathcal{I}_{(S+S)E}^{(t)}=S(\Lambda_t\otimes\Lambda_t [\rho_{S+S}])-\frac{1}{4}\ \sum_{\mathclap{i,
			j=1,2}} S(e^{t\mathcal{L}_i}\otimes e^{t\mathcal{L}_j}[\rho_{S+S}]).
\end{equation}
Notice that, unlike in the unitary case, each of the entropies in the second term now grows in time due to the joint unital dissipative evolution of $\mathcal{A}_S$ and $\mathcal{A}_E^{(0)}$ that mixes them. 

\begin{figure}[t]
	\centering
	\includegraphics[width=0.47\linewidth]{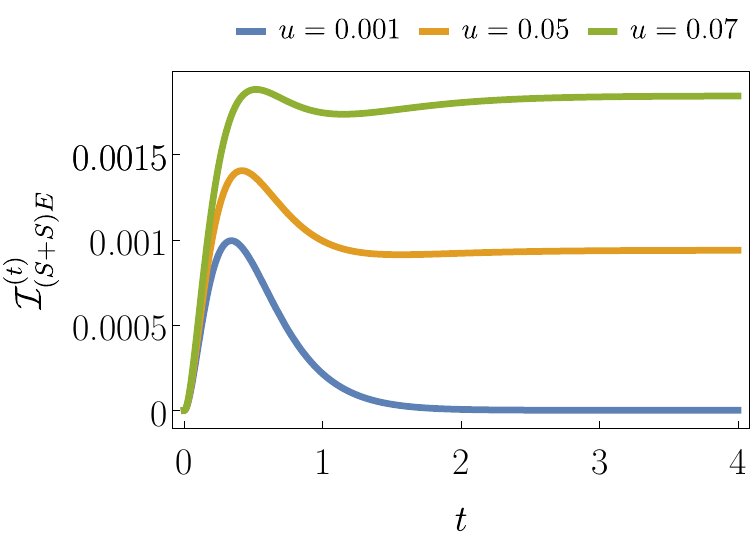}\qquad
	\includegraphics[width=0.47\linewidth]{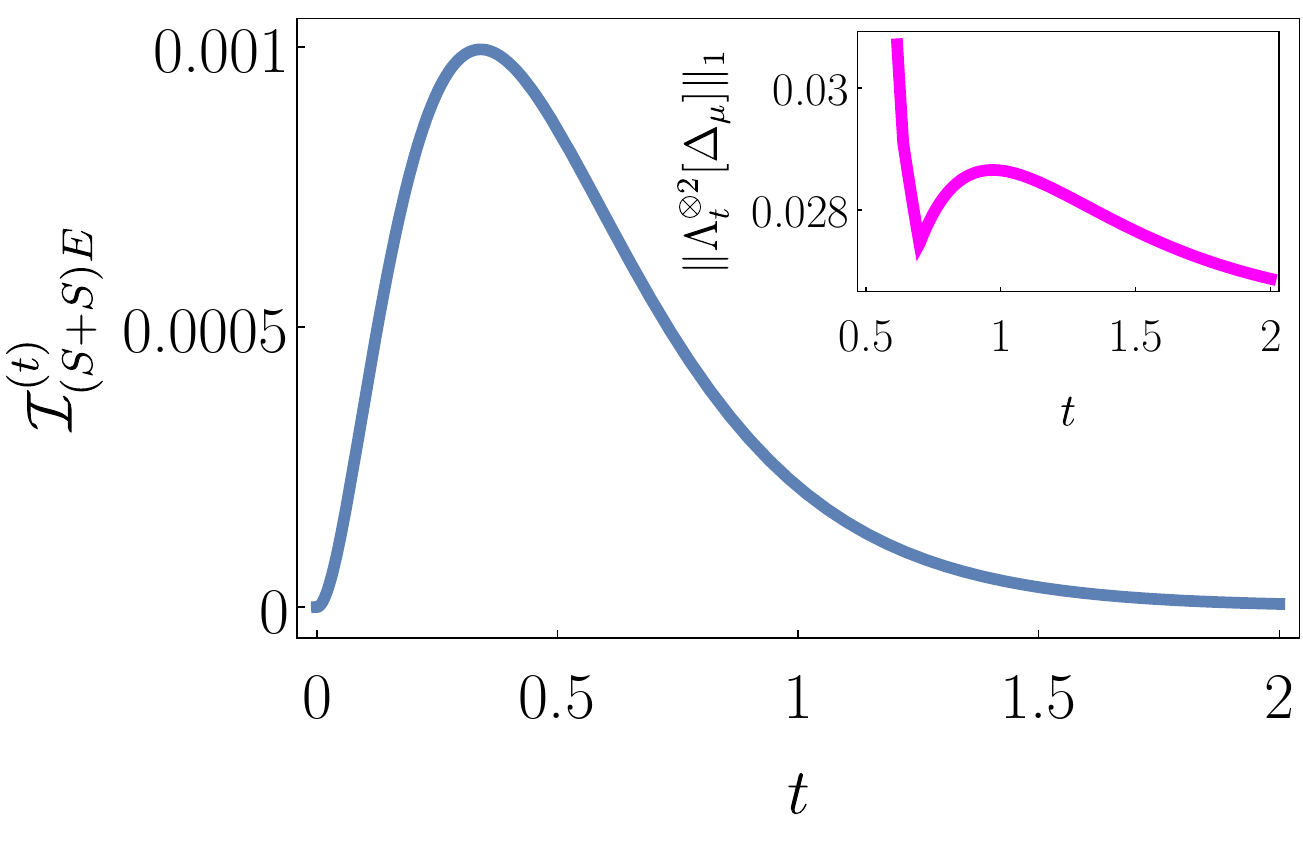}
	\caption{
		System-chain mutual information for the state $\rho_X^{(1)}$ with fixed $\mu_1=\mu_2=\nu=1/4, v=i/8$ at different values of real $u$.
		For $u=0.001$, the behaviour is compared with that of  the trace norm of the Helstrom matrix between $\Lambda_t\otimes\Lambda_t[\rho_X^{(1)}]$ and  $\Lambda_t\otimes\Lambda_t[\rho_X^{(3)}]$, with bias $\mu=0.52$, $\rho_X^{(3)}$ being in the form~\eqref{Xshape} w.r.t.the computational basis and defined by parameters $\mu_1'=\mu_2'=1/2,\nu'=0, u'=1/8,v'=0$. One easily sees that the marginals of $\rho_X^{(1)}$ are the maximally mixed state $\mathds{1}_2/2$. 
	}
	\label{fig:mutualvsinternal}
\end{figure}

We study $\mathcal{I}_{(S+S)E}^{(t)}$ picking $\rho_{S+S}$ of $X$-shape with respect to the eigenvectors of the matrix $\sigma_1\otimes\sigma_1$:
\begin{align}\label{Xshape}
	\rho_X^{(1)}&=\begin{pmatrix}
		\mu_1 & 0 & 0 & u \\
		0 	  & \nu & v & 0 \\
		0	  & \bar{v} & 1-(\mu_1+\mu_2+\nu) &0	\\
		\bar{u} & 0 & 0 & \mu_2
	\end{pmatrix}.
\end{align}
In Appendix~\ref{app:Xx}, its decomposition in terms of the Pauli matrix tensor products $\{\sigma_i\otimes\sigma_j\}_{ij}$ is reported, from which the time-evolving states entering \eqref{qmi_continuous} can be easily inferred.
In Figure~\ref{fig:mutualvsinternal}, we display the system-chain mutual information when the system is initialized in a state of the class~\eqref{Xshape}, which displays a growth and collapse of correlations. We also compare such behaviour with that of $\normt{\Lambda_t\otimes\Lambda_t[\Delta_\mu(\rho_X^{(1)},\rho_X^{(3)})]}$, where $\rho_X^{(3)}$ has $X$ shape in the computational basis.

Thus, the system-chain correlations can undergo a decrease for a certain time interval, despite the stationarity of the environment.
\begin{remark}
	The information lost by the system and subjected to BFI is generally thought to be stored either in system-environment correlations or in changes of the environmental state (notice that in our Example, the environment is stationary~\eqref{envstationarity})~\cite{ColloquiumBreuer,HolevoSkew}.
	In Fig.\ref{fig:mutualvsinternal} the mutual information $\mathcal{I}_{SE}^{(t)}$ of~\eqref{qmi_continuous} is plotted for X states with $\mu_{1,2}=\nu=
	1/4$.  As for the maximally entangled state state $P_2^+$ considered in~\eqref{computationdeltaQMI}, these states have maximally mixed marginals. For a state $\rho_{S+S}$ with maximally mixed marginals, using trace preservation and factorization, one shows that 
	$$
	\Tr_{1(2)}(\Lambda_t\otimes\Lambda_t[\rho_{S+S}])=\Lambda_t[\Tr_{1(2)}(\rho_{S+S})]=\frac{\mathds{1}_2}{2}\ .
	$$ 
	Similarly, one checks that the one-qubit local density matrix~\eqref{localSE}, obtained by tracing over one of the two open systems 
	together with its own environment, reduces to
	\begin{equation}\label{uncorrelatedmarginals}
		\Omega_{S[-a,b]}^{(t)}=\frac{\mathds{1}_2}{2}\otimes\rho_E^{[-a,b]}\implies \mathcal{I}_{SE}^{(t)}=0.
	\end{equation}
	For such states, the bipartite correlations have a non-monotonic behaviour in time, while the qubit-chain marginals are uncorrelated at all times due to~\eqref{uncorrelatedmarginals}. Thus, in such case, the information is temporarily  stored non-locally in the system-environment correlations.
\end{remark}

\section{Quantum signature of SBFI.}   
SBFI is undoubtedly a memory effect with no classical counterpart, despite it might arise from the coupling to a classical collisional environment. 
The reason is that positivity and complete positivity coincide for mappings on commutative algebras. 
To illustrate this in more detail recall that, in a commutative setting, the Helstrom matrix takes the form
\begin{equation}
	\Delta_\mu(\rho_S,\sigma_S)=\sum_i (\mu p_i-(1-\mu) q_i) P_i\,,
\end{equation}
$p_i, q_i$ being, respectively, the eigenvalues of $\rho_S$ and $\sigma_S$ and $P_i$ their common eigenprojectors. Thus, the Helstrom distinguishability reduces to the $\ell_1$-norm of the vector $\norm{\mu\ket{p}-(1-\mu)\ket{q}}_{\ell_1}$, with $\norm{\ket{x}}_{\ell_1}=\sum_{i=1}^d \abs{x_i}$.
In the case of a classical bipartite system, consider a real vector $\ket{x}=\sum_{ij}x_{ij}\ket{i}\otimes\ket{j}\in \mathbb{R}^{d}\otimes\mathbb{R}^{d}$ evolving into $\ket{x_t} =T(t)\otimes T(t) \ket{x}$, under the action of a continuous-time  P-divisible stochastic process $T(t)$, such that for all $t\ge s\ge0$,  $T(t)=T(t,s)T(s)$, with $T(t,s)$ a stochastic matrix, $T_{ik}(t,s)\ge0$ and $\sum_i T_{ik}(t,s)=1$.  
Under such dynamics, the $\ell_1$-norm of a time-evolving vector $\ket{x}=\{x_{ij}\}\in\mathbb{R}^d\times\mathbb{R}^d$  cannot increase in time,%; namely,
\begin{equation}
	\partial_t\norm{T(t)\otimes T(t)\ket{x}}_{\ell_1}\le0\,,\quad \forall \ket{x} \in \mathbb{R}^d\times\mathbb{R}^d\,.
\end{equation}
Indeed,
\begin{align*}
	\norm{\ket{x_{t}}}_{\ell_1}=\sum_{ij}\left\vert \sum_{kl}T_{ik}(t,s) \, T_{jl}(t,s) x_{kl}(s)\right\vert  &\le \sum_{kl} \sum_{ij}|T_{ik}(t,s)|\, |T_{jl}(t,s)|\, \abs{x_{kl}(s)}\\
	&\le \sum_{kl} \abs{x_{kl}(s)}=\norm{\ket{x_{s}}}_{\ell_1}\ ,
\end{align*}
for all $t\ge s\ge0$. For quantum systems, 
 as we have seen, the phenomenon of SBFI is witnessed by the quantity
\begin{align}\label{witness_SBFI}
	\Delta D_\mu{(t+\tau,t)}&:=\normt{\Lambda_{t+\tau}\otimes\Lambda_{t+\tau}[\Delta_\mu(\rho_{S+S},\sigma_{S+S})]}\nonumber\\&\hskip.5cm
	-\normt{\Lambda_{t}\otimes\Lambda_t[\Delta_\mu(\rho_{S+S},\sigma_{S+S})]}\,, 
\end{align}
assuming a strictly positive value at some $t,\tau>0$, that is by revivals of the bipartite Helstrom distinguishability.
The quantum character of such a memory effect can be assessed by the following measure of the quantum correlations present in the Helstrom ensemble $\mathcal{E}_H(t)=\{(\mu; \rho_{S+S}(t)),(1-\mu;\sigma_{S+S}(t))\}$.
The quantumness of a single-party ensemble $\mathcal{E}=\{(\mu_i,\rho_i)\}$ has been identified with the possibility of simultaneously diagonalizing it \cite{Groisman2007_ArXiv,LuoEquantumness_2010}; equivalently, if the ensemble is encoded into a quantum-classical state $\chi^\mathcal{E}=\sum_{i} \mu_i\rho_i\otimes \ketbra{i}$, one can measure the ensemble quantumness in terms of the quantum correlations as left-sided quantum discord in $\chi^\mathcal{E}$ \cite{LuoEquantumness_2010,LuoEQuantumness_2011}. Among the variety of available discord measures \cite{AdessoCorrelations2016}, we shall consider the so-called measurement induced geometric measure of quantum correlations defined in the trace norm by
\begin{align*}
	\mathcal{Q}_{\{\mathbb{P}\}}(\rho):=\min_{\mathbb{P}}D(\rho ,   \, \mathbb{P}\otimes \mathrm{id}[\rho]) \,,
\end{align*}
where $\mathbb{P}[X]=\sum_i P_i X P_i$ is a projective measurement associated to $=\{P_i\}_i,  P_i=\ketbra{i}$ an orthonormal set of rank-1 projectors.
If $\mathcal{E}$  is an ensemble of bipartite states, one rather focuses on finding a simultaneous diagonalization on a set of rank-1 projections of the type $\{P_i^1\otimes P_j^2\}_{ij}$. Accordingly, in \cite{PianiQuantumnessCorrelations2014}  the following measure of bipartite ``ensemble quantumness of correlations'' was introduced:
\begin{equation}\label{ensemble_quantumnesscorrelations}
	\mathcal{Q}_{\{\mathbb{P}^1\otimes\mathbb{P}^2\}}(\chi^{\mathcal{E}})=\min_{\mathbb{P}^1\otimes \mathbb{P}^2}\sum_i \mu_i \, D(\rho_i ,\,\mathbb{P}^1\otimes \mathbb{P}^2
	[\rho_i] )\,,
\end{equation}
with $\chi^\mathcal{E}$ now encoding the bipartite ensemble by means of an additional classical register.

The next result connects the bipartite quantumness of correlations of the Helstrom ensemble, as defined by~\eqref{ensemble_quantumnesscorrelations}, to the quantity~\eqref{witness_SBFI} exposing SBFI.
\begin{proposition}\label{prop:quantumnessSBFI_bound}
	Given a dynamics $\Lambda_t\otimes\Lambda_t$ with $\Lambda_t$ P-divisible, the variation of the Helstrom distinguishability   $\Delta D_\mu{(t+\Delta t,t)}$ can be bounded as follows
	\begin{align}\label{bound_Ensemblequantumness}
		\Delta D_\mu{(t+\tau,t)}\le2\,\normd{\Lambda_{t+{\tau},t}}^2 \  \,\mathcal{Q}_{\{\mathbb{P}^1\otimes \mathbb{P}^2\}}({\chi^{\mathcal{E}_H}}(t))\,,
	\end{align}
	where $\chi^{\mathcal{E}_H}(t)=\mu \rho_{S+S}(t)\otimes \ketbra{0}+ (1-\mu) \sigma_{S+S}(t)\otimes \ketbra{1}$, while $\normd{\,\cdot\,}$ denotes the diamond norm of a map.%\footnote{The diamond norm of $\Lambda:\Md{d}\to \Md{d'}$ is given by $\max \{\normt{\Lambda\otimes\mathrm{id}_d[X]}:\normt{X}\le 1\}$} (see also~\cite{WatrousQI}). 
\end{proposition}

\begin{proof}
	Let us fix $\{\ket{p_\alpha}\}_\alpha$ with $\ket{p_\alpha}=|{p_i^1}\rangle\otimes |{p_j^2}\rangle$, $\{|{p_i^1}\rangle\}_i$, $\{|{p_j^2}\rangle\}_j$ being arbitrary local orthonormal basis, from which one has a corresponding  orthonormal set of rank-1 projectors
	$
	\{P_i^1\otimes P_j^2\}_{ij}$. 
	Accordingly, a completely decohering map with respect to such basis is described by a (bi)local projective measurement:
	$$	
	{ \mathbb{P}_1\otimes\mathbb{P}_2}[X]=\sum_{ij}P_i^1\otimes P_j^2  \,X P_i^1\otimes P_j^2\ .
	$$
	Then, for $t>s>0$ both in discrete and continuous time, considering the Helstrom matrix  at time $t$, 
	$\Lambda_t\otimes\Lambda_t[\Delta_\mu(\rho_{S+S},\sigma_{S+S})]$, via the triangle inequality and the contractivity of $\Lambda_{t,s}$ and ${\mathbb{P}_1\otimes\mathbb{P}_2}$, one estimates
	\begin{align}\nonumber
		\normt{(\Lambda_{t,s}\otimes\Lambda_{t,s})\circ ({\mathbb{P}_1\otimes\mathbb{P}_2})[\Delta_\mu(s)]}
		&\le \sum_{ij} 	\big| \delta_{\mu}^{ij}(s) \big|\, \normt{\Lambda_{t,s}[P_i^1]} 		\,	\big\|\Lambda_{t,s}[P_j^2]\big\|_1 \\
		&\le \normt{{\mathbb{P}_1\otimes\mathbb{P}_2}[\Delta_\mu(s)]} \le {\normt{\Delta_\mu(s)}}\ ,
		\label{first_estimate_proof}
	\end{align}
	where $\delta_\mu ^{ij}(s):=\langle p_i^1|\langle p_j^2|\Delta_\mu(s)|p_i^1 \rangle |p_j^2\rangle$.
	Consider the induced trace norm and the diamond norm of $\Lambda: \Md{d}\to \Md{d'}$ \cite{WatrousQI},
	$$
	\normt{\Lambda}=\max \{ \normt{\Lambda[X]}: {\normt{X}\le1}\}\ ,\quad 
	\normd{\Lambda}=\normt{\Lambda\otimes \mathrm{id}_d}\ .
	$$
	Then, the variation of the Helstrom matrix, 
	$$
	{\Delta {D}_\mu(t,s):=\normt{\Lambda_{t,s}\otimes\Lambda_{t,s}[\Delta_\mu(s)]}-\normt{\Delta_\mu(s)}}
	$$ 
	can be upper-bounded as follows
	%\begin{widetext}
	\begin{align*}
		\Delta {D}_\mu(t,s)&= \,\big\|{\Lambda_{t,s}\otimes\Lambda_{t,s}[\Delta_\mu(s)]}\big\|_1-\normt{\Delta_\mu(s)}	\\
		&=   \, \big\|\mu \,\Lambda_{t,s}\otimes\Lambda_{t,s}\big[\rho_{S+S}(s)-{\mathbb{P}_1\otimes\mathbb{P}_2}[\rho_{S+S}(s)]\big]  \\& \hskip0.5cm -(1-\mu)\Lambda_{t,s}\otimes \Lambda_{t,s}\big[\sigma_{S+S}(s)-{\mathbb{P}_1\otimes\mathbb{P}_2}[\sigma_{S+S}(s)]\big] +\Lambda_{t,s}\otimes\Lambda_{t,s}\circ{\mathbb{P}_1\otimes\mathbb{P}_2}[\Delta_\mu(s)] \big\|_1 \\& \hskip0.5cm -\normt{\Delta_{\mu}(s)} \\
		&\le \ \mu \,\normt{\Lambda_{t,s}\otimes\Lambda_{t,s}\big[\rho_{S+S}(s)-{\mathbb{P}_1\otimes\mathbb{P}_2}[\rho_{S+S}(s)]\big]}\\ &\hskip0.5cm +(1-\mu)\,\normt{\Lambda_{t,s}\otimes\Lambda_{t,s}\big[\sigma_{S+S}(s)-{\mathbb{P}_1\otimes\mathbb{P}_2}[\sigma_{S+S}(s)]\big]}  \\& \hskip0.5cm	+\normt{\Lambda_{t,s}\otimes\Lambda_{t,s}\circ{\mathbb{P}_1\otimes\mathbb{P}_2}[\Delta_\mu(s)]}
		-\normt{\Delta_{\mu}(s)}\,.
	\end{align*}
	%\end{widetext}
	Using \eqref{first_estimate_proof} and the fact that  $\normt{\Lambda\otimes \Lambda}\le \normd{\Lambda}^2 $, we have
	\begin{align*}\label{latter_proof}
		\Delta {D}_\mu(t,s) \le &\,\normd{\Lambda_{t,s}}^2\,  \,\big(\mu \normt{\rho_{S+S}(s)-{\mathbb{P}_1\otimes\mathbb{P}_2}[\rho_{S+S}(s)]} \\ &\hskip4cm +(1-\mu)\normt{\sigma_{S+S}(s)-{\mathbb{P}_1\otimes\mathbb{P}_2}[\sigma_{S+S}(s)]}\big)\,.
	\end{align*}
	Since $\mathbb{P}_{1,2}$ are arbitrary, one can tighten the latter inequality by minimizing over the projective measurements. 
	One then finally obtains the following upper-bound for $\Delta {D}_\mu(t,s)$,
	\begin{equation*}
		\Delta {D}_\mu(t,s) \ \le \  2\,\normd{\Lambda_{t,s}}^{2}  \, \, \mathcal{Q}_{\{\mathbb{P}^1\otimes\mathbb{P}^2\}}(\chi^\mathcal{E}(s))\,,
	\end{equation*}
	where the quantumness of the Helstrom ensemble $\mathcal{E}_H(s)=\{(\mu;\rho_{S+S}(s)),(1-\mu;\sigma_{S+S}(s))\}$, encoded  in  the quantum-classical state $$\chi^\mathcal{E}(s)=\mu\, \rho_{S+S}(s)\ketbra{0}+(1-\mu)\,\sigma_{S+S}(s)\otimes\ketbra{1}\,,$$
	is measured by the (left-sided) quantum correlations of 
	$\chi^{\mathcal{E}_H}(s)$~\cite{PianiQuantumnessCorrelations2014}:
	\begin{align*}
		\mathcal{Q}_{\{\mathbb{P}^1\otimes\mathbb{P}^2\}}(\chi^{\mathcal{E}_H}(s))&=\frac{1}{2} \min_{\mathbb{P}^1\otimes\mathbb{P}^2} \big(\mu \normt{\rho_{S+S}(s)-{\mathbb{P}^1\otimes\mathbb{P}^2}[\rho_{S+S}(s)]}\\
		&\hskip4cm	+(1-\mu)\normt{\sigma_{S+S}(s)-{\mathbb{P}^1\otimes\mathbb{P}^2}[\sigma_{S+S}(s)]}\big)\,.\qedhere
	\end{align*} 
\end{proof}

%The proof % of \eqref{bound_Ensemblequantumness} 
%can be found  in Appendix~\ref{app:bound}. 

\begin{remark}
	If SBFI triggers at time $t$, i.e. $\Delta D_\mu{(t+\tau,t)}>0$, then the quantumness of correlations of the ensemble $\mathcal{E}_H(t)=\{(\mu,\rho_{S+S}(t));(1-\mu, \sigma_{S+S}(t))\}$ has to be strictly positive, that is, the state $\chi^{\mathcal{E}_H}(t)$ has to have a non zero quantum discord. 
	In this sense, the Helstrom ensemble quantumness is a ``precursor'' of non-Markovianity~\cite{Precursors}. In particular, $\mathcal{Q}_{\{\mathbb{P}^1\otimes \mathbb{P}^2\}}({\chi^{\mathcal{E}_H}}(t))>0$ does not imply that the states are entangled (as similarly noted in~\cite{BuscemiDatta,acin2017constructive}; a simple construction in the Pauli case for a quantum ensemble triggering SBFI but not involving entanglement is reported in Appendix~\ref{app:noentanglementconstruction}).
\end{remark}

\color{black}

\section{Conclusions}  

In this work we studied the SBFI in an open  system of two qubits, each coupled to a classical Markov chain. 
The assumptions made in the treatment of the environment and the interaction allowed for a full analytical description of the system dynamics and system-environment correlations. Notice that there is a structural difficulty, both analytically and from the point of view of a microscopic derivation, to devise dynamics that are P-divisible but not CP-divisible. This reflects the lack of a general characterization of positive maps versus completely positive ones. Despite these general obstructions, the proposed model is sufficiently rich to provide a dynamics with a neat microscopic origin of its degree of divisibility, and able to display the SBFI effect.

Both in the discrete and continuous-time regimes, we investigated the emergence of bipartite memory effects by means of the system-chain mutual information  of local density matrices obtained through an algebraic approach. Growths and collapses of correlations have been detected for both unitary and dissipative collisions: in the former case, the mutual information is simply  the system's entropy up to a constant, while in the latter case it has the form of a Jensen-Shannon divergence. 
Despite the ongoing debate regarding the physical nature of Backflow of Information, especially in such kind of classical environments, the non-monotonicity of the aforementioned quantities provides a clear-cut physical interpretation in terms of system-environment correlations.
Interestingly, despite information might be stored in and released through classical correlations, SBFI has no classical counterpart; however,  the quantum resource needed to trigger it is only the quantumness of the Helstrom ensemble but not entanglement in its states.
%\textcolor{blue}{however, entanglement is not needed in order to trigger it. Rather, the necessary quantum resource is only the quantumness of the Helstrom ensemble.}

\paragraph{Acknowledgements}
F.B. and G.N. acknowledge financial support from PNRR MUR project PE0000023-NQSTI.

%\clearpage
\appendix
\begin{appendices}
\counterwithin*{equation}{section}
\renewcommand\theequation{\thesection\arabic{equation}}

	\section{Reduced dynamics}
	\label{app:reduceddyn}
	
	As seen in the main text, tensor product elements of the local algebra $\mathcal{A}^{[-a,b]}_E$ supported by the interval of integers $-a\leq j\leq b$ are denoted by
	means of the multi-indices $\bs{i}_{[-a,b]}=i_{-a}i_{-a+1}\cdots i_b$ as follows:
	$$
	A^{\otimes[-a,b]}_{\bs{i}_{[-a,b]}}=A^{(-a)}_{i_{-a}}\otimes A^{(-a+1)}_{i_{-a+1}}\otimes\cdots\otimes A^{(b)}_{i_b}=\bigotimes_{k=-a}^bA^{(k)}_{i_k}\ ,
	$$
	where the upper index in $A^{(k)}_{i_k}$ indicates the site $k$ at which the operator $A_{i_k}$ is located.
	
	The collisional dynamics $\Phi_n=(\Theta\circ\Phi)^n$ comprises 1) the right shift $\Theta$ on $\mathcal{A}_E$ such that
	$$
	\Theta\Big[A^{[-a,b]}_{\bs{i}_{[-a,b]}}\Big]=A^{[-a+1,b+1]}_{\bs{i}_{[-a,b]}}\ ,
	$$ 
	and 2) the CPU map on the bipartite algebra $\mathcal{A}_S\otimes\mathcal{A}$ of system $S$ and chain ancilla at site $0$ defined by:
	\begin{equation}
		\label{aux0}
		\Phi\Big[O_S\otimes A_{i_0}^{(0)}\Big]=\sum_{k=0}^{d-1}\phi_{k}\left[O_S\right] \otimes \Pi^{(0)}_k A^{(0)}_{i_0} \Pi^{(0)}_k\ ,\quad 
		A_{i_0}\in\mathcal{A}\quad\hbox{at site $0$}\ ,
	\end{equation} 
	with $\phi_k$ completely positive, unital maps on the system algebra $\mathcal{A}_S$. When extended to the whole algebra $\mathcal{A}_S\otimes\mathcal{A}_E$, $\Phi$ yields 
	\begin{equation}
		\label{sppl00}
		\Phi_1\Big[O_S\otimes A^{\otimes[-a,-1]}_{\bs{i}_{[-a,-1]}}\otimes A^{(0)}_{i_0}\otimes A^{\otimes[1,b]}_{\bs{i}_{[1,b]}}\Big]=
		\sum_{k=0}^{d-1}\phi_{k}[O_S] \otimes A^{\otimes[-a+1,0]}_{\bs{i}_{[-a,-1]}}\otimes\Pi^{(1)}_k A^{(1)}_{i_0}\Pi^{(1)}_k\otimes 
		A^{\otimes[2,b+1]}_{\bs{i}_{[1,b]}}\ .
	\end{equation}
	Iterating the action of $\Theta\circ\Phi$ one gets
	\begin{align}\nonumber
		\Phi_n\Big[O_S\otimes A^{\otimes[-a,b]}_{\bs{i}_{[-a,b]}}\Big]\ &=\
		\sum_{\bs{k}_{[1,n]}} \phi_{\bs{k}_{[1,n]}}[O_S]\otimes A^{\otimes[-a+n,0]}_{\bs{i}_{[-a,-n]}}\otimes\Pi_{k_1}^{(1)}A^{(1)}_{i_{-n+1}}\Pi_{k_1}^{(1)}\otimes\cdots \\
		& \hskip6cm\otimes
		\Pi_{k_n}^{(n)}A^{(n)}_{i_0}\Pi_{k_n}^{(n)}\otimes A_{i_{[1,b]}}^{\otimes[n+1,b+n]}\ ,	\label{sppl0}
	\end{align}
	where $\bs{k}_{[a,b]}$ denotes the multi-index $k_ak_{a+1}\ldots k_b$ and $\phi_{\bs{k}_{[1,n]}}\equiv\phi_{k_1}\circ\phi_{k_2}\circ\cdots\circ\phi_{k_n}$.
	
	\subsection{System $S$ reduced dynamics}
	\label{app:reddynS}
	
	The reduced dynamics $\Lambda_n$ of the states of  the open system $S$ at discrete time $n$ in~\eqref{reduceddynamicsn}
	is obtained through (see~\eqref{marginal1}), 
	$$
	{\rm Tr}\Big(\Omega_S^{(n)}O_S\Big)=\omega_{SE}\circ\Phi_n\Big(O_S\otimes\mathds{1}_E\Big)\ ;
	$$
	namely by restricting the compound state $\omega_{SE}\circ\Phi_n$ to the system $S$ algebra $\mathcal{A}_S\otimes\mathds{1}_E$.
	Using~\eqref{sppl0} one gets
	\begin{equation}
		\label{sppl1}
		\Phi_n[O_S\otimes\mathds{1}_E]=
		\sum_{\bs{k}_{[1,n]}} \phi_{\bs{k}_{[1,n]}}[O_S]\otimes \bigotimes_{j=1}^n \Pi_{k_j}^{(j)}\ .
	\end{equation}
	Let us consider an initial factorized state $\omega_S\otimes\omega_E$ where the system state $\omega_S$ is represented by a density matrix $\rho_S$, while the restriction of the environment state $\omega_E$ to the algebra spanned by the orthogonal projections 
	$ \Pi_{\bs{k}_{[1,n]}}\equiv\bigotimes_{j=1}^n \Pi_{k_j}^{(j)}$ gives rise to the density matrix $\sum_{\bs{k}_{[1,n]}}p_{\bs{k}_{[1,n]}}\Pi_{\bs{k}_{[1,n]}}$. 
	Then,
	$$
	\omega_S\otimes\omega_{E}(\Phi_n[O_S\otimes\mathds{1}_E])=\Tr(\rho_{S}\sum_{\bs{k}_{[1,n]}}p_{\bs{k}_{[1,n]}}\phi_{\bs{k}_{[1,n]}}[O_S])
	=\Tr(\sum_{\bs{k}_{[1,n]}}p_{\bs{k}_{[1,n]}}^{\phantom{\ddagger}}\phi_{\bs{k}_{[1,n]}}^\ddagger[\rho_S]O_S)\ ,
	$$
	where $\Phi_{\bs{k}_{[1,n]}}^\ddagger=\phi_{k_n}^\ddagger\circ\cdots\circ\phi_{k_1}^\ddagger$ with $\phi_{k_i}^\ddagger$ the dual map of $\phi_{k_i}$. Hence, in the Schrödinger picture, the dynamical map reads
	\begin{equation}\label{reduced_app}
		\Lambda_n=\sum_{\bs{i}_{[1,n]}}p_{\bs{i}_{[1,n]}}^{\phantom{\ddagger}}\phi_{\bs{i}_{[1,n]}}^\ddagger\,.
	\end{equation}

	\subsection{Environment $E$ reduced dynamics}
	\label{app:reddynE}
	
	The single site operators $A_{i}$ belong to the commutative algebra generated by the orthogonal projectors $\Pi^{(k)}_j$; then, 
	$\sum_{k=0}^{d-1}\Pi_kA_i\Pi_k=A_i$.
	Therefore, due to the assumed unitality of the CP maps $\phi_k$, from~\eqref{sppl0} it follows that
	\begin{align}
		\nonumber
		\Phi_n\Big[\mathds{1}_S\otimes A^{\otimes[-a,b]}_{\bs{i}_{[-a,b]}}\Big]&=
		\sum_{\bs{k}_{[1,n]}} \mathds{1}_S\otimes A^{\otimes[-a+n,0]}_{\bs{i}_{[-a,-n]}}\otimes\Pi_{k_1}^{(1)}A^{(1)}_{i_{-n+1}}\Pi_{k_1}^{(1)}\otimes\cdots \\
		&\nonumber\hskip7cm\otimes
		\Pi_{k_n}^{(n)}A^{(n)}_{i_0}\Pi_{k_n}^{(n)}\otimes A_{\bs{i}_{[1,b]}}^{\otimes[n+1,b+n]}   \\ 
		&\nonumber=\mathds{1}_S\otimes A^{\otimes[-a+n,0]}_{\bs{i}_{[-a,-n]}}\otimes A^{(1)}_{i_{-n+1}}\otimes\cdots\otimes
		A^{(n)}_{i_0}\otimes A_{\bs{i}_{[1,b]}}^{\otimes[n+1,b+n]}
		\\&=\mathds{1}_S\otimes\Theta^n\Big[A^{\otimes[-a,b]}_{\bs{i}_{[-a,b]}}\Big]\ .	\label{sppl2}	
	\end{align}
	Since the environment state is shift-invariant by construction, it follows that the environment state is stationary:
	$$
	\omega_{SE}\circ\Phi_n\Big(\mathds{1}_S\otimes \mathcal{A}_E\Big)=\omega_{SE}\Big(\mathds{1}_S\otimes \mathcal{A}_E\Big)=\omega_E\Big(\mathcal{A}_E\Big)\ .
	$$

	\section{$\Lambda_n$ in the unitary case}
	\label{app:unitary}
	
	The unitary case correspond to choosing $\varphi=-1$ in~\eqref{Paulieig}. Then, only $j=n-2$ contributes to the sum in~\eqref{spectrumrecurrence} so that:
	\begin{equation}
		\label{recurr1_proof}
		\lambda_n^{(\ell)}=(1-2(p+r))\,\lambda_{n-1}^{(\ell)}\,+4\, p\, \Delta \,\lambda_{n-2}^{(\ell)}\ ,\quad\ell=1,2\ .
	\end{equation}
	The general solutions of \eqref{recurr1_proof} can be found with the ansatz $\lambda_n^{(\ell)}=x\lambda_{n-1}^{(\ell)}$ for all $n\geq 2$, by means of
	the roots $x^\pm$ of~\cite{concrete_tetrahedron}
	\begin{equation}
		\label{aux2}
		P(x)=x^2-\alpha\, x\,+\,4p\Delta\ ,\qquad \alpha=1-2(p+r) \ .
	\end{equation}
	
	The general solution will thus have the form
	$\displaystyle
	\lambda_{n}^{(\ell)}=c_+\, x_+^n + c_- \,x_-^n$, with the 
	constants $c_\pm$ fixed by the initial conditions $\lambda_0^{(\ell)}=1$ and $\lambda_1^{(\ell)}=\alpha$. The eigenvalues $\lambda_n^{(1,2,3)}$ then read
	\begin{align}
		\label{aux3a}
		\lambda_n^{(\ell)}&=	\frac{\beta+\alpha}{ 2\beta}	\left(\frac{\alpha+\beta}{ 2}	\right)^n	+\frac{\beta-\alpha}{ 2\beta} \left(\frac{\alpha-\beta}{2}\right)^n=:\lambda_n\ ,\quad\beta={\sqrt{\alpha^2+16\,p\Delta}\ ,}\quad\ell=1,2\ ,\\
		\label{aux3b}	\lambda_{n}^{(3)}&=\,(1-4p)^n\, .
	\end{align}
	From the multiplicative action of the Pauli maps $\Lambda_n$ on the Pauli matrices, 
	one deduces that $\Lambda_n$ is a convex combination of two discrete-time semigroups:
	\begin{eqnarray}
		\label{decomp1}
		\Lambda_n&=& \frac{\beta+\alpha}{ 2\beta} \Psi_+^n + \frac{\beta-\alpha}{ 2\beta}\  \Psi_-^n\ ,\qquad 
		\Psi_\pm[X]=\sum_{i=0}^3 \psi_\pm^{(i)}\, \Tr(\sigma_i X)\, \sigma_i\ ,\qquad \hbox{where}\\ 
		\label{decomp1bis}
		\psi_{\pm}^{(1,2)}&=&\frac{\alpha\pm\beta}{2}\ ,\qquad
		\psi_\pm^{(0)}=1\ ,\ \psi_\pm^{(3)}=1-4p\ .
	\end{eqnarray}
	It will be sufficient to consider the case $\alpha>0$, namely $r<1/2-p$. If $p\ne1/4$, then $\lambda^{(j)}_n$ and $\Lambda_n$.
	We can thus compute the intertwining maps $\Lambda_{n,n-1}=\Lambda_n^{\phantom{-1}}\circ\Lambda_{n-1}^{-1}$ between two subsequent collisions. Setting $\displaystyle \gamma:=\frac{\beta+\alpha}{2}>\frac{\beta-\alpha}{2}=:\delta>0$, these maps are of Pauli type with eigenvalues
	\begin{equation}
		\label{contractivity}
		\lambda_{n,n-1}:=\lambda_{n,n-1}^{(1)}=\lambda_{n,n-1}^{(2)}=\frac{\lambda_n}{\lambda_{n-1}}=\frac{\gamma^{n+1}+(-1)^n\, \delta^{n+1}}{\gamma^n+(-1)^{n-1}\,\delta^n}\,,\qquad 	\lambda_{n,n-1}^{(3)}=1-4p	\ .
	\end{equation}
	The P-divisibility of the discrete family of Pauli maps $\Lambda_n$, that is the contractivity of the intertwining maps $\Lambda_{n.n-1}$, is equivalent to asking that
	%\begin{equation}
	$\left|\lambda_{n,n-1}^{(i)}\right|\le1$, $i=1,2,3$.
	%\end{equation}
	In order to show this, we first prove that
	\begin{equation}
		\label{contraux}
		\lambda_{2,1}>\lambda_{n,n-1}\qquad \forall n>2\ .
	\end{equation}
	To see this, let $[0,1]\ni x \equiv \delta/\gamma$. For even $n=2k>2$, 
	\begin{equation}
		\lambda_{n,n-1}=\gamma\,  \frac{1+x^{n+1}}{1-x^n}\,,
	\end{equation} 
	monotonically decreases with $n$. Instead, for odd $n=2k+1>2$, 
	\begin{equation}
		\lambda_{n,n-1}=\gamma\,  \frac{1-x^{n+1}}{1+x^n}\,,
	\end{equation} 
	increases with $n$; nevertheless, 
	\begin{equation}
		\frac{\lambda_{n,n-1}}{\lambda_{2,1}}=\,  \frac{1-x^{n+1}}{1+x^n}\frac{1-x^2}{1+x^3}\le 1\,.
	\end{equation}
	Notice that $p_0=1-r-2p\geq 0$ and $r\geq 0$ imply $0\leq p\leq 1/2$, so that
	$\abs{\lambda_{n,n-1}^{(3)}}=\abs{1-4p}\le1$. From~\eqref{contractivity} and the previous discussion, one checks when $\abs{\lambda_{2,1}}\le1$:
	\begin{align}\label{conditionPd}
		0\le\lambda_{2,1}=1-2(p+r)+\frac{4\,\Delta\, p}{1-2(p+r)}\le1\iff \frac{\Delta}{\alpha}\le \frac{r}{2p}+\frac{1}{2}\ ,
	\end{align}
	where $\alpha$ has been defined in~\eqref{aux2}.
	
	Instead, the conditions for the complete positivity of $\Lambda_{n,n-1}$ are obtained by asking for the positivity of the eigenvalues of the $4\times 4$ Choi matrix $\Lambda_{n,n-1}\otimes\mathrm{id}[P_+^2]$, where 
	$P_+^2$ projects onto the totally symmetric vector 
	$\displaystyle 
	\ket{\Psi_+^2}=\frac{1}{\sqrt{2}}(\ket{00}+\ket{11})$.
	The eigenvalues are easily computed to be: $E_1(n)=p$, twice degenerate and 
	\begin{align}
		E_2(n)=&\frac{1}{4}(1+\lambda_{n,n-1}^{(3)}+2\lambda_{n,n-1} )=\frac{1-2p}{2}+\frac{\lambda_{n,n-1}}{2}\ ,
		\label{choieig_1}\\ 
		E_3(n)=&\frac{1}{4}(1+\lambda_{n,n-1}^{(3)}-2\lambda_{n,n-1})=\frac{1-2p}{2}-\frac{\lambda_{n,n-1}}{2}\ .
		\label{choieig_2}
	\end{align}
	From $0\leq p\leq 1/2$ and $\lambda_{n,n-1}\geq 0$ it follows that 
	$E_1(n)\geq 0$. Further,~\eqref{contraux} implies $E_3(n)\geq E_3(2)$; then the positivity of $E_3(n)$ is ensured by
	\begin{align}
		\label{conditionCPd} 
		1+\lambda_{2,1}^{(3)}-2\lambda_{2,1}=4r-\frac{8\,p\,\Delta}{\alpha}\ge0 \iff \frac{\Delta}{\alpha}\le\frac{r}{2p}\,.
	\end{align}
	We now consider the positivity of $\Lambda_{n,n-1}\otimes\Lambda_{n,n-1}$. 
	Since $\Lambda_{n,n-1}$ are Pauli maps, then $\Lambda_{n,n-1}\otimes\Lambda_{n,n-1}$ is  positive if and only if $\Lambda_{n,n-1}^2$
	is completely positive~\cite{FilippovMagadov}, that is if and only if the Choi matrix 
	$\Lambda_{n,n-1}^2\otimes\mathds{1}[P^2_+]\geq 0$. Recasting
	$$
	P^{(2)}_+=\frac{1}{4}\Big(\mathds{1}\otimes\mathds{1}+\sigma_1\otimes\sigma_1-\sigma_2\otimes\sigma_2+\sigma_3\otimes\sigma_3\Big)=
	\frac{1}{4}
	\begin{pmatrix}
		\mathds{1}+\sigma_3&\sigma_1+i\sigma_2\cr
		\sigma_1-i\sigma_2&\mathds{1}-\sigma_3
	\end{pmatrix}
	$$
	yields
	%the action $\Lambda[\mathds{1}]=\mathds{1}$
	%and $\Lambda[\sigma_k]=\ell_k\sigma_k$, $k=1,2,3$, of a unital Pauli map yields
	%\begin{equation}
	%\label{symprojLambda}
	%\Lambda\otimes\Lambda[P^{(2)}_+]
	%=\Lambda^2\otimes\mathds{1}[P^{(2)}_+]
	%=
	%\frac{1}{4}
	%\begin{pmatrix}
	%1+\ell_3^2&0&0&\ell_1^2+\ell_2^2\cr0&1-\ell_3^2&\ell_1^2-\ell_2^2&0\cr
	%0&\ell_1^2-\ell_2^2&1-\ell_3^2&0\cr\ell_1^2+\ell_2^2&0&0&1+\ell_3^2
	%\end{pmatrix}\ .
	%\end{equation}
	\begin{align}
		\label{symprojLambda}
		\Lambda_{n,n-1}\otimes\Lambda_{n,n-1}[P^{(2)}_+]
		&=\Lambda_{n,n-1}^2\otimes\mathds{1}[P^{(2)}_+]\nonumber\\
		&\hskip-2cm=
		\frac{1}{4}
		\begin{pmatrix}
			1+(\lambda_{n,n-1}^{(3)})^2&0&0&2\lambda_{n,n-1}^2\cr
			0&1-(\lambda_{n,n-1}^{(3)})^2&0&0\cr
			0&0&1-(\lambda_{n,n-1}^3)^2&0\cr
			2\lambda_{n,n-1}^2&0&0&1+(\lambda^{(3)}_{n,n-1})^2
		\end{pmatrix}\ .
	\end{align}
	%The positivity of the Choi-matrix is thus equivalent to $0\leq \ell_3^2\leq 1$ and
	%$1\pm\ell^2_3\geq \vert\ell_1^2\pm\ell_2^2\vert$.
	Then, %setting $\ell_{k}=\lambda^{(k)}_{n,n-1}$ as in~\eqref{contractivity},
	from $0\leq p\leq1/2$ it follows that $\Lambda_{n,n-1}\otimes\Lambda_{n,n-1}$ is 
	completely positive  iff 
	\begin{equation}
		\label{aux1}
		1+(1-4p)^2-2\big(\lambda_{n,n-1}\big)^2\ge0\ .
	\end{equation}
	Moreover, since $\lambda_{n,n-1}\geq\lambda_{2,1}\geq 0$, 
	we get the inequality
	\begin{equation}\label{conditiontensorpower}
		4\, p^2 \left(\frac{\Delta}{\alpha}\right)^2+2\,p\,\alpha\left(\frac{\Delta}{\alpha}\right)-(p^2+r \, p_0)\le0\ ,
	\end{equation}
	where we recall that $p_0=1-2p-r$. 
	Equation~\eqref{conditiontensorpower} then gives the condition for P-divisibility of $\Lambda_n\otimes\Lambda_n$,
	\begin{equation}\label{conditionLambda2}
		\frac{\Delta}{\alpha}\le\frac{r}{2p}+\frac{1}{2}-\frac{1-\sqrt{1-4p(1-2p)}}{4p}\,.
	\end{equation}
	Clearly,  $\eqref{conditionCPd}\implies\eqref{conditionLambda2}\implies\eqref{conditionPd}$. On the other hand, $\eqref{conditionLambda2}\centernot\implies\eqref{conditionCPd}$. This is in contrast to the case of continuous-time dynamics with a time-local generator. Indeed, 
	as proved in Theorem 1 of~\cite{BenattiChrusFil}, positive $\Lambda_{t,s}\otimes\Lambda_{t,s}$ for all $t\ge s\ge 0$ are possible if and only if $\Lambda_{t,s}$ are completely positive.

	To see explicitly how the environmental correlations relate to the lack of BFI for one qubit and super-activation of BFI for two qubits, let us consider $r=0$ and $p \ll 1$ and let $\Delta=Q\, p$. From \eqref{aux3a} and \eqref{aux3b}, one sees that
	\begin{align}
		\lambda_{n,n-1}&=1-2p 
		+4Q p^2+O(p^3)\,,%+8Q\varepsilon^3 +o (\varepsilon^3) \,,
		\qquad \lambda_{n,n-1}^{(3)}=1-4p\,,
	\end{align}
	for all $n\ge2$. 
	Notice that the eigenvalues of $X=X^\dagger=x_0+\sum_{i=1}^3 x_i\sigma_i$ in $\Md{2}$ are $x_0\pm\norm{\bs{x}}$. Then, $\normt{X}=2x_0={\rm Tr}X$ if $x_0\geq \norm{\bs{x}}$, otherwise $\normt{X}=2\norm{\bs{x}}$.
	Let us assume $\Tr(X)=2x_0\ge0$ and set $Y=\Lambda_{n,n-1}[X]$,  its eigenvalues being $x_0\pm \norm{\bs{y}}$, with $\bs{y}=(\lambda_{n,n-1} x_1,\lambda_{n,n-1} x_2,\lambda_{n,n-1}^{(3)}x_3)$. 
	Thus, $\normt{Y}=\normt{X}=2x_0$ if $x_0>\norm{\bs{y}}$, otherwise $\normt{Y}=2\norm{\bs{y}}$.
	Then, expanding up to the second order in $p$ one finds
	\begin{align}
		\label{smallnorm}
		\norm{\bs{y}}^2= \norm{\bs{x}}^2 -2p\,\Big(2(x_1^2+x_2^2)+x_3^2)\Big)+4p^2\Big(Q(x_1^2+x_2^2)+x_3^2\Big)+O(p^3)\ .
	\end{align}
	Therefore, for $x_0\leq\norm{\bs{y}}$, $x_0\leq\norm{\bs{x}}$ so that $\normt{Y}\leq\norm{\bs{x}}=\normt{X}$ and contractivity ensues.
	%\begin{align}\label{smallnorm}
	%\norm{\bs{y}}= \norm{\bs{x}} -2\,\frac{x_1^2+x_2^2+2x_3^2}{\norm{\bs{x}}} p+2 (x_1^2+x_2^2)\frac{x_3^2+2Q\norm{\bs{x}}^2}{\norm{\bs{x}}^3} p^2 +O(p^3)\,.
	%\end{align}
	Indeed that a $Q$-dependent, positive contribution in~\eqref{smallnorm} only appears at second order in $p$ and is dominated by a strictly negative contribution, thus preventing BFI for one qubit.  Also, notice that up to second order in $p$ there is no dependence on the successive discrete-time steps $n$ and $n-1$.
	
	Instead, let us consider the case of two qubits and consider the trace norm of 
	$Z:=\Lambda_{n,n-1}\otimes\Lambda_{n,n-1}[P_2^+]$ in the same small $p$ regime. 
	From~\eqref{symprojLambda} one sees that the eigenvalues of $Z$ are $1-(\lambda^{(3)})^2\geq 0$ twice degenerate and
	$$
	1+(\lambda^{(3)})^2+2\lambda_{n,n-1}^2\geq 0\ ,\qquad
	1+(\lambda^{(3)})^2-2\lambda_{n,n-1}^2\ .
	$$
	If the latter is positive it follows that $\normt{Z}=1=\normt{P_2^+}$; otherwise, if $2\lambda^2_{n,n-1}>1+(\lambda^{(3)}_{n,n-1})^2$, which for small $p$ occurs whenever $Q>1/2$,  
	$$
	\normt{Z}=2\left(1-(\lambda^{(3)})^2\right)+4\lambda^2_{n.n-1}\simeq 1+4 p^2(2Q-1)
	$$
	becomes larger than $1$ for $Q>1/2$.
	Therefore, unlike for a single qubit, for two qubits the leading correction  is a term of order $2$ in $p$; this becomes positive for sufficiently correlated  
	sites in the Markov chain environment in which case $\Lambda_{n,n-1}\otimes\Lambda_{n,n-1}$ ceases to be contractive.
	%An  eigenvalue $2 p^2 (1 - 2 Q)$. If $Q>1/2$,
	%$$
	%\normt{\Lambda_{2,1}^{\otimes{2}}[P_2^+]}=1+4 p^2(2Q-1)>1=\normt{P_2^+}\,.
	%$$

	\section{Local system-chain density matrices and mutual information}\label{app:localdm}
	\label{app:mutinf}
	
	Let us consider again the local  algebra $\mathcal{A}_E^{[-a,b]}$ supported by the integers $0\leq a\leq j\leq b$ whose elements are linear combinations of tensor products $A_{\bs{i}_{[-a,b]}}^{\otimes[-a,b]}$.
	Each single-site operator belongs to the commutative algebra $\mathcal{A}=D_d(\mathbb{C})$ generated by the orthogonal projections $\Pi_j$, $0\leq j\leq d-1$ and is thus of the form
	$A^{(k)}_{i_k}=\sum_{\ell_k=0}^{d-1}a_{i_k}^{\ell_k}\Pi^{(k)}_{\ell_k}$. Then,
	$$
	A_{\bs{i}_{[-a,b]}}^{\otimes[-a,b]}=\sum_{\bs{\ell}_{[-a,b]}} a^{\bs{\ell}_{[-a,b]}}_{\bs{i}_{[-a,b]}}\Pi^{[-a,b]}_{\bs{\ell}_{[-a,b]}}\ ,\quad 
	\Pi^{[-a,b]}_{\bs{\ell}_{[-a,b]}}\equiv\bigotimes_{k=-a}^b \Pi^{(k)}_{\ell_k}\ ,\quad a^{\bs{\ell}_{[-a,b]}}_{\bs{i}_{[1-a,b]}}\equiv\prod_{k=-a}^ba_{i_k}^{\ell_k}\ .
	$$
	The dynamics~\eqref{sppl0} thus gives
	\begin{equation}
		\label{observablecases}
		\Phi_n\left[O_S\otimes	A_{\bs{i}_{[-a,b]}}^{\otimes[-a,b]}\right]
		=\left\{\begin{matrix}
			
			\sum_{\bs{\ell}_{[-a,b]}} \,  a^{\bs{\ell}_{[-a,b]}}_{\bs{i}_{[-a,b]}} \phi_{\bs{\ell}_{[-n+1,0]}}[O_S]\otimes
			\Pi_{\bs{\ell}_{[-a,b]}}^{[-a+n,n+b]}&\ldots&0\leq n\leq a\\
			\\
			\sum_{\bs{\ell}_{[-n+1,b]}} \,  a^{\bs{\ell}_{[-a,b]}}_{\bs{i}_{[-a,b]}} \phi_{\bs{\ell}_{[-n+1,0]}}[O_S]\otimes
			\Pi_{\bs{\ell}_{[-n+1,b]}}^{[1,n+b]}&\ldots&n> a
		\end{matrix}
		\right.\ ,
	\end{equation} 
	where $\phi_{\bs{\ell}_{[-n+1,0]}}\equiv \phi_{\ell_{-n+1}}\circ\cdots\circ\phi_{\ell_0}$.
	
	Let us now consider the discrete-time evolution of local density matrices that is obtained by duality:
	$$
	\omega_S\otimes\omega_E\circ \Phi_n\left[O_S\otimes A_E^{[-a,b]}\right]=\Tr\left(\Omega_{S[-a,b]}^{(n)}\, O_S\otimes A_E^{[-a,b]}\right)\,.
	$$
	Using the shift invariance of the environment state $\omega_E$ one gets:
	\begin{align*}
		\omega_S\otimes\omega_E\circ\Phi_n\left[O_S\otimes	A_{\bs{i}_{[-a,b]}}^{\otimes[-a,b]}\right] & =
		\left\{\begin{matrix}
			\sum_{\bs{\ell}_{[-a,b]}} p_{\bs{\ell}_{[-a,b]}}
			a^{\bs{\ell}_{[-a,b]}}_{\bs{i}_{[-a, b]}} \,  \Tr(\phi^\ddagger_{\bs{\ell}_{[-n+1,0]}}[\rho_S]\,O_S)&\dots \quad n\leq a\\
			\\
			\sum_{\bs{\ell}_{[-n+1,b]}} p_{\bs{\ell}_{[-n+1,b]}}
			a^{\bs{\ell}_{[-a,b]}}_{\bs{i}_{[-a, b]}} \,  \Tr(\phi^\ddagger_{\bs{\ell}_{[-n+1,0]}}[\rho_S]\,O_S) &\dots \quad n>a
		\end{matrix}\right. .
	\end{align*}
	Therefore, the local density matrices at discrete time-step $n$, $\Omega_{S [-a,b]}^{(n)}$, read
	\begin{equation}
		\label{OmegaSE}
		\Omega_{S [-a,b]}^{(n)}=\left\{\begin{matrix}
			\sum_{\bs{\ell}_{[-a,b]}}  p_{\bs{\ell}_{[-a,b]}}\phi^\ddagger_{\bs{\ell}_{[-n+1,0]}}[\rho_S]\otimes 
			\Pi_{\bs{\ell}_{[-a,b]}}^{[-a,b]}&\ldots&n\leq a\\
			\\
			\sum_{\bs{\ell}_{[-n+1,b]}}  p_{\bs{\ell}_{[-n+1,b]}}\phi^\ddagger_{\bs{\ell}_{[-n+1,0]}}[\rho_S]\otimes 
			\Pi_{\bs{\ell}_{[-n+1,b]}}^{[-n+1,b]}&\ldots& n>a
		\end{matrix}
		\right.\ .
	\end{equation}
	To quantify the system-chain correlations, we compute the mutual information
	\begin{equation}
		\label{app:mutinf1}
		\mathcal{I}\left(\Omega_{S [-a,b]}^{(n)}\right)=S\left(\Omega_{S }^{(n)}\right)+S \left(\Omega_{ [-a,b]}^{(n)}\right)- S\left(\Omega_{S [-a,b]}^{(n)}\right)\ ;
	\end{equation}
	relative to the evolved local density matrices $\Omega_{S [-a,b]}^{(n)}$~\eqref{OmegaSE} and their marginals $\Omega_S^{(n)}$, respectively 
	$\Omega_{ [-a,b]}^{(n)}$ that are obtained by performing the trace over $\mathcal{A}_S$, respectively $\mathcal{A}_E$.
	Using the notation in~\eqref{chainstate}, they read 
	\begin{eqnarray}
		\label{app:mutinf2a}
		\Omega_{S }^{(n)}&=&
		\sum_{\bs{\ell}_{[-n+1,0]}}  p_{\bs{\ell}_{[-n+1,0]}}\phi^\ddagger_{\bs{\ell}_{[-n+1,0]}}[\rho_S]\ ,\\
		\label{app:mutinf2b}
		\Omega_{ [-a,b]}^{(n)}&=&
		\left\{
		\begin{matrix}
			\sum_{\bs{\ell}_{[-a,b]}}  p_{\bs{\ell}_{[-a,b]}}\,\Pi_{\bs{\ell}_{[-a,b]}}^{[-a,b]}=\rho_E^{[-a,b]}&\ldots&n\geq a\\
			\\
			\sum_{\bs{\ell}_{[-n+1,b]}}  p_{\bs{\ell}_{[-n+1,b]}}\,\Pi_{\bs{\ell}_{[-n+1,b]}}^{[1-n,b]}=\rho_E^{[-n+1,b]}&\ldots&n\geq a
		\end{matrix}
		\right.
		\ .
	\end{eqnarray}
	Notice that~\eqref{app:mutinf2a} follows since
	$$
	a\geq n \Rightarrow \sum_{\bs{\ell}_{[-a, -n]};\bs{\ell}_{[1,b]}}\,p_{\bs{\ell}_{[-a,b]}}= p_{\bs{\ell}_{[-n+1,0]}}\ .
	$$
	Furthermore, by relabelling the indices in the right-hand side of~\eqref{app:mutinf2a}, one obtains $\Omega_{S }^{(n)}=\Lambda_n[\rho_S]$ with $\Lambda_n$ as in~\eqref{reduced_app}.

	Since the contributing operators in~\eqref{OmegaSE} are all orthogonal, one gets
	\begin{equation}
		\label{globent}
		S\left(\Omega_{S [-a,b]}^{(n)}\right)=\left\{
		\begin{matrix}
			S\left(\rho_E^{[-a,b]}\right)+\sum_{\bs{\ell}_{[-n+1,0]}}p_{\bs{\ell}_{[-n+1,0]}}\,S\left(\phi^\ddagger_{\bs{\ell}_{[-n+1,0]}}\right)&\ldots&n\leq a\\
			\\
			S\left(\rho_E^{[-n+1,b]}\right)+\sum_{ \bs{\ell}_{[-n+1,0]}} p_{\bs{\ell}_{[-n+1,0]}}\, S\left(\phi^\ddagger_{\bs{\ell}_{[-n+1,0]}}[\rho_S]\right)&\ldots&n>a
		\end{matrix}
		\right.
		\ .
	\end{equation}
	Therefore, by relabeling the summation indices,
	% The latter further simplifies by using the compatibility and stationarity properties of the Markov chain,
	% \begin{align*}
		% 	S\left(\Omega_{S [-a+1,b]}^{(n)}\right)&=S\left(\Omega_{ [-a+1,b]}^{(n)}\right)\  +\sum_{ \bs{i}^{(n)}} \, p_{\bs{i}^{(n)}}  \   S\left(\Phi^\ddagger_{\bs{i}^{(n)}}[\rho_S]\right)\,.
		% \end{align*}
	% Similarly, one sees that
	% $	\Omega_{S}^{(n)}=\Lambda_n[\rho_S]$, 
	% so that 
	the mutual information simplifies to
	\begin{equation}\label{mutualI}
		\mathcal{I}\left(\Omega_{S [-a+1,b]}^{(n)}\right)=S(\Lambda_n[\rho_S])-\sum_{ \bs{\ell}_{[1,n]}} p_{\bs{\ell}^{[1,n]}}  
		\ S\left(\Phi^\ddagger_{\bs{\ell}_{[1,n]}}[\rho_S]\right)\equiv\mathcal{I}_{SE}^{(n)}\,.
	\end{equation}
	Notice that the right-hand side of~\eqref{mutualI} only depends on $n$ and not on the specific local sub-algebra $\mathcal{A}_E^{[-a,b]}$
	%a,b$. %$a=0$ and $b=0$.
	%For the case $n> a$,  an observable $\widetilde{O}_E^{[-a+1,b]}$,  can be   embedded into  $O_E^{[-n+1,b]}=\mathds{1}^{[-n+1,-a]}\otimes \widetilde{O}_E^{[-m+1,b]}$.
	% If $n>a$, instead,
	%\begin{align*}
	%\mathcal{I}\left(\widetilde{\Omega}_{S [-a+1,b]}^{(n)}\right)&= S\bigg(\widetilde{\Omega}_{S [-n+1,b]}^{(n)}\bigg\| \widetilde{\Omega}_{S }^{(n)}\otimes\widetilde{\Omega}_{ [-n+1,b]}^{(n)}\bigg)	\\
	%&=S\bigg(
	% \Tr_{[-n+1,-a]}\left(\Omega_{S [-n+1,b]}^{(n)}\right)\bigg\|\Tr_{[-n+1,-a]}\left( \Omega_{S }^{(n)}\otimes\Omega_{ [-n+1,b]}^{(n)}\right)\bigg)	\\
	%& \le 
	%	S\bigg(\Omega_{S [-n+1,b]}^{(n)}\bigg\| \Omega_{S }^{(n)}\otimes\Omega_{ [-n+1,b]}^{(n)}\bigg)=\mathcal{I}_{SE}^{(n)}\,,
	% \end{align*}
% where in the last line the monotonicity of the relative entropy under the partial trace was exploited. Thus, the quantity $\mathcal{I}_{SE}^{(n)}$ in \eqref{mutualI} %should be regarded as the maximal system-chain quantum mutual information. 
Analogously, for two qubits evolving in the same collisional environment, the mutual information relative to an initial state $\omega_{S+S}\otimes\omega_E$ where $\omega_{S+S}$ is an expectation corresponding to a two-qubit state $\rho_{S+S}$, one similarly derive 
\begin{equation*}
	\mathcal{I}_{SE}^{(n)}:=S(\Lambda_n\otimes\Lambda_n[\rho_{S+S}])-\sum_{\bs{\ell}_{[1,n]},\bs{k}_{[1,n]}} p_{\bs{\ell}_{[1,n]}} p_{\bs{k}_{[1,n]}} \ S\left(\phi^\ddagger_{\bs{\ell}_{[1,n]}}\otimes\phi^\ddagger_{\bs{k}_{[1,n]}}[\rho_{S+S}]\right)\,.
\end{equation*}

\section{Stroboscopic limit}
\label{app:strobo}

Let us consider Pauli maps as in~\eqref{paulidiss} of the form $\phi_k=e^{\tau\mathcal{L}_k}$ and $\varphi=e^{-2\gamma\tau}$. This choice corresponds to the case in which the system, identified by $\mathcal{A}_S$, and the chain ancilla at site $(0)$, described by $\mathcal{A}_E^{(0)}$, are dissipatively coupled for a time $\tau$ through the following GKLS generator, 
\begin{align}
	\label{app:GKSL)}
	\mathbb{L}[O_S\otimes O_E^{(0)}]&=\gamma\sum_{i=0}^3 \left((\sigma_i\otimes\Pi_i)  \, O_S\otimes O_E^{(0)} \, (\sigma_i\otimes\Pi_i)-\frac{1}{2}\left\{\mathds{1}_2\otimes\Pi_i ,O_S\otimes O_E^{(0)}\right\}\right)\ ,
\end{align}
which satisfies $\mathbb{L}[O_S\otimes \Pi_j]=\mathcal{L}_j[O_S]\otimes\Pi_j$.
The reduced dynamics will be of the Pauli type, with spectrum $\lambda_n^{(i)}$ obeying the recurrences~\eqref{spectrumrecurrence} and~\eqref{spectrumrecurrence1}.
In the so-called stroboscopic limit typical of collision models, one takes  
\begin{equation}\label{strobo}
	\tau\to0\,,\qquad n\to \infty, \qquad n\tau\to t \ ,
\end{equation} 
and expands~\eqref{spectrumrecurrence1} at first order in $\tau$ obtaining 
\begin{equation}
	\label{Paulieigcont}
	\frac{\lambda_{n}^{(3)}-\lambda_{n-1}^{(3)}}{\tau}=-2\gamma \lambda_{n-1}^{(3)} \implies \dot\lambda_{t}^{(3)}=-2\gamma\lambda_t^{(3)}\Rightarrow
	\lambda_t^{(3)}=e^{-2\gamma t}\ .
\end{equation}
%is straightforward and 
On the other hand, denoting by $\lambda_n$ the other two equal Pauli eigenvalues and expanding~\eqref{spectrumrecurrence} up to order $\tau$ %$(g\tau)^2$ 
yield the following finite-difference equation:
\begin{equation}\label{intermediate}
	\frac{\lambda_n-\lambda_{n-1}}{\tau}=-2(p+r)\gamma\lambda_{n-1}+2p\,\gamma^2 \,\sum_{j=0}^{n-2}\tau \,(2\Delta)^{n-j-1}(1-\gamma \tau)^{n-j-2} \lambda_j\,.
\end{equation}
Choosing
$\displaystyle\Delta=\frac{e^{-\kappa\tau}}{2}\,,$
the stroboscopic limit~\eqref{strobo} and the constraints~\eqref{prob-const} yield $p\to 1/2, r\to 0$ and turn \eqref{intermediate} into the integro-differential equation
\begin{equation}
	\dot{\lambda}_t=-\gamma\,\lambda_t+\gamma^2 \,\int_{0}^t\dd{s} e^{-(\kappa+\gamma)(t-s)}\lambda_s\,.
\end{equation}
The latter is readily solvable through its Laplace transform $\widetilde{\lambda_z}=\int_0^{+\infty}\dd{t}e^{-zt}\lambda_t$, with the initial condition $\lambda_{t=0}=1$, yielding:
\begin{equation}\label{laplace}
	\widetilde{\lambda_z}=\frac{z+\kappa+\gamma}{z^2+z(\kappa+2\gamma)+\kappa\gamma}\quad\hbox{with simple poles at}\quad
	z_{\pm}=\frac{-(\kappa+2\gamma)\pm\sqrt{\kappa^2+4\gamma^2}}{2}\le0\ .	
\end{equation}
Therefore, for $a\ge z_+$, one gets
\begin{equation}
	\lambda_t=\frac{1}{2\pi i} \int_{a-i\infty}^{a+i\infty} \dd{z}e^{zt}\widetilde{\lambda}_z %\frac{z+\kappa+\gamma}{z^2+z(k+2\gamma)+k\gamma}
	%=\frac{z_{+} +\kappa+\gamma}{z_+-z_-} {e^{z_-\,t}}+\frac{z_{-} +\kappa+\gamma}{z_--z_+} {e^{z_+\,t}}\label{laplacesol1}\\
	=\	e^{-(\gamma +\frac{\kappa}{2}) t } \bigg[\cosh \left(\frac{1}{2} t \sqrt{\kappa^2+4 \gamma ^2}\right) +\frac{\kappa \sinh \left(\frac{1}{2} t \sqrt{\kappa^2+4 \gamma ^2}\right)}{\sqrt{\kappa^2+4
			\gamma ^2}}\bigg]\,\label{laplacesol2}.
\end{equation}
By inspection of the Choi matrix of $\Lambda_t$, namely $\Lambda_t\otimes\mathrm{id}_2[P_2^+]$, one realizes that complete positivity requires
\begin{equation}
	1+\lambda_t^{(3)}-2\lambda_t\ge0\,,
\end{equation}
which is checked to be always satisfied. 

\begin{remark}
	\label{rem:appE}
	
	From~\eqref{Paulieigcont} and~ \eqref{laplacesol2}, one derives that the dynamical map $\Lambda_t$ can be written as a convex composition of two-semigroups,
	\begin{equation}
		\Lambda_t=a e^{t \mathcal{L}_-}+(1-a)e^{t \mathcal{L}_+} \,, \quad 0\leq a=\frac{1}{2}+\frac{\kappa}{2\sqrt{\kappa^2+4\gamma^2}}\leq 1\ .
		%\frac{z_{+} +\kappa+\gamma}{z_+-z_-}\in[0,1]\ .
	\end{equation}
	Notice that only $e^{t \mathcal{L}_-}$ is completely positive, while $e^{t \mathcal{L}_+}$ is only positive. Nevertheless, $\Lambda_t$ is always completely positive and P-divisible. 
	Indeed, the contractivity of the the Pauli intertwiners $\Lambda_{t,s}$, 
	as discussed in Appendix~\eqref{app:unitary} for the discrete-time case,
	is equivalent to requiring that the Pauli eigenvalues are monotonically decreasing functions of time,
	\begin{equation}
		\dot{\lambda}_t\le0\,,\qquad\dot{\lambda}_t^{(3)}\le0\,.
	\end{equation}
	This is verified since the Pauli spectrum evolves according to
	\begin{equation}
		\dot\lambda_t = -\Gamma_t\, \lambda_t\,,\qquad 	\dot\lambda_t^{(3)} = -\Gamma^{(3)}_t\, \lambda_t^{(3)}\,,
	\end{equation}
	where $\Gamma_t^{(3)}=2\gamma\ge0$ and 
	\begin{equation*}
		\Gamma_t=\gamma-\frac{2 \gamma ^2}{\sqrt{\kappa^2+4 \gamma ^2} \coth \left(\frac{1}{2} t \sqrt{\kappa^2+4 \gamma ^2}\right)+\kappa}\,.
	\end{equation*}
	The case $\kappa=0$ has already been discussed in the main text, while the positivity of $\Gamma_t$ for $\kappa>0$ is equivalent to
	\begin{equation}
		1+\sqrt{1+\left(\frac{2\gamma}{\kappa}\right)^2}\coth(  \frac{t}{2}\sqrt{\kappa^2+4\gamma^2})\ge \frac{2\gamma}{\kappa}\,,
	\end{equation}
	which is clearly verified since $\coth(x)\ge1$ for $x\ge0$ .
\end{remark}

\section{Details about the Mutual information}\label{app:Xx}

In the case of a unitary coupling between system and collisional environment, the variation of the system-chain mutual information reduces to the variation 
of the von Neumann entropy in discrete time as in~\eqref{mutualfinitedifference}. 
As considered in the main text, the choice~\eqref{choicepar} together with $p=1/4+\epsilon$ yield the following Pauli eigenvalues at discrete-time steps $1$, respectively $2$:
$\lambda_1=\lambda_2=\frac{1}{2}-2\epsilon$, respectively $\lambda_1^{(3)}= -4\epsilon$, $\lambda_2^{(3)}=16\epsilon^2$.

Since $\epsilon$ is taken as a small perturbative parameter, it follows that the  intertwiner $\Lambda_{2,1}$ is a positive map. Indeed, the corresponding Pauli eigenvalues satisfy 
$$
\lambda_{2,1}=\frac{\lambda_2}{\lambda_1}=1\ ,\quad \abs{\lambda_{2,1}^{(3)}}=\abs{\frac{\lambda^{(3)}_2}{\lambda^{(3)}_1}}=4\epsilon<1\ .
$$
Then, consider the first two time-step dynamics of two-qubit totally symmetric state $P_2^+$:
%under the second tensor power after the first and the second collision,
\begin{align}
	\label{examplemutual}
	\Lambda_1\otimes\Lambda_1[P_2^+]&=\begin{pmatrix}
		\frac{1}{4}+4\epsilon^2 & \cdot & \cdot & \frac{1}{8}(1-4\epsilon)^2 \\
		\cdot & \frac{1}{4}-4\epsilon^2  & \cdot & \cdot \\
		\cdot &  \cdot & \frac{1}{4}-4\epsilon^2& \cdot \\
		\frac{1}{8}(1-4\epsilon)^2 & \cdot & \cdot & \frac{1}{4}+4\epsilon^2
	\end{pmatrix}
	\\
	\label{examplemutual2}
	\Lambda_2\otimes\Lambda_2[P_2^+]&=\begin{pmatrix}
		\frac{1}{4}+64\epsilon^4 & \cdot & \cdot & \frac{1}{8}(1-4\epsilon)^2 \\
		\cdot & \frac{1}{4}-64\epsilon^4  & \cdot & \cdot \\
		\cdot &  \cdot & \frac{1}{4}-64\epsilon^4& \cdot \\
		\frac{1}{8}(1-4\epsilon)^2 & \cdot & \cdot & \frac{1}{4}+64\epsilon^4
	\end{pmatrix}.
\end{align}
By evaluating the spectrum of the two states and expanding the  von Neumann entropies of the two above states~\eqref{examplemutual} up to second order in $\epsilon$, one gets
\begin{align*}
	S(\Lambda_1\otimes\Lambda_1[P_2^+])&=\frac{20\log(2)-3\log(3)}{8}+\log(3)\,\epsilon+\left(8\log(2)-6\log(3)-\frac{16}{3}\right)\epsilon^2+O(\epsilon^3)\,,\\
	S(\Lambda_2\otimes\Lambda_2[P_2^+])&=\frac{20\log(2)-3\log(3)}{8}+\log(3)\,\epsilon-\left(2\log(3)+\frac{16}{3}\right)\epsilon^2+O(\epsilon^3)\,.
\end{align*}
Their difference coincides with the variation of the system-chain correlations and is given, up to order $\epsilon^2$, by
$$
\Delta \mathcal{I}_{SE}^{(2,1)}(P_2^+)=S(\Lambda_2\otimes\Lambda_2[P_2^+])-S(\Lambda_1\otimes\Lambda_1[P_2^+])=-4\log(\frac{4}{3})\epsilon^2+O(\epsilon^3)<0\,.
$$
Let us now compute the quantum mutual information for the case $p=1/2$, $\Delta=1/2$, corresponding to master equation rates
$\gamma_t=1$, $\gamma_t^{(3)}=-\tanh(t)$ and Pauli eigenvalues
$\lambda_t=e^{-t}\cosh(t)$, $\lambda_t^{(3)}=e^{-2t}$.
Notice that, with these choices, the stochastic matrix $T$ in~\eqref{tmatrix} takes a particularly simple form:
\begin{equation}\label{specialT}
	T=\frac{1}{2}\begin{pmatrix}
		0 & 0 & 0 & 0 \\
		1 & 2 &	0 & 1 \\
		1 & 0 & 2 & 1 \\
		0 & 0 & 0 & 0
	\end{pmatrix},
\end{equation}
so that the only non-zero probabilities $p_{\bs{i}_{[1,n]}}$ correspond to the sequences 
$\bs{i}_{[1,n]}=111\dots\equiv\bs{1}$ and 
$\bs{i}_{[1,n]}=222\dots\equiv \bs{2}$.
Accordingly, only two CPTP unital semigroups $\phi_{\bs{i}_{[1,n]}}$ %corresponding to above two  sequences 
contribute in \eqref{reduced_dyn},
\begin{equation}
	\Lambda_t=\frac{\phi_{\bs{1}}+\phi_{\bs{2}}}{2}\,,
\end{equation}
\begin{comment}
	, the only relevant emerging CPTP unital semigroups $\phi_{\bs{i}_{[1,n]}}$ in \eqref{label} correspond to the constant sequences $\bs{i}_{[1,n]}=111\dots\equiv\bs{1}$, 
	$\bs{i}_{[1,n]}=222\dots\equiv \bs{2}$ 
\end{comment}
with equal, time-independent weights $p_{\bs{1}}=p_{\bs{2}}=1/2$. In the continuous-time limit, $\phi_{\bs{1}}$ and $\phi_{\bs{2}}$  are the Pauli maps defined by
\begin{equation}
	\begin{split}
		\phi_{\bs{1}}[\sigma_1]&=1\,,\\
		\phi_{\bs{2}}[\sigma_1]&=e^{-2t}\,,\\
	\end{split}
	\qquad
	\begin{split}
		\phi_{\bs{1}}[\sigma_2]&=e^{-2t}\,,\\
		\phi_{\bs{2}}[\sigma_2]&=1\,,\\
	\end{split}
	\qquad
	\begin{split}
		\phi_{\bs{1}}[\sigma_3]&=e^{-2t}\,,\\
		\phi_{\bs{2}}[\sigma_3]&=e^{-2t}\,.\\
	\end{split}
\end{equation}
Notice that for other choices of $T$, in the stroboscopic limit, the weights $p_{\bs{i}_{[1,n]}}$ would generally become functions of time as well. 
In the special case of~\eqref{specialT}, the mutual information as function of $t$ reads 
\begin{equation}\label{mutualinfor2sequences}
	\mathcal{I}_{SE}^{(t)}(\rho_S)=S(\Lambda_t\otimes \Lambda_t)[\rho_{S+S}])-\frac{1}{4}\sum_{i,j=\bs{1},\bs{2}} S(\phi_{i}\otimes \phi_{j}[\rho_{S+S}])\,.
\end{equation}
The non-monotonic behaviour of the system-chain mutual information~\eqref{mutualinfor2sequences} has been inspected numerically by means of the following family of $X$ states,
\begin{align}\label{Xx}
	\rho_X^{(1)}&=\frac{1}{4}\big\{\mathds{1}_4-(1-2(\mu_1+\nu))\sigma_1\otimes\mathds{1}_2+(1-2(\mu_2+\nu))\mathds{1}_2\otimes\sigma_1\nonumber\\&\hskip0.5cm-(1-2(\mu_1+\mu_2))\sigma_1\otimes\sigma_1-2\Re(u-v)\sigma_2\otimes\sigma_2	\nonumber\\&\hskip0.5cm	+2\Re(u+v)\sigma_3\otimes\sigma_3+2\Im(u+v)\sigma_2\otimes\sigma_3+2\Im(u-v)\sigma_3\otimes\sigma_2\big\}\,,
\end{align}
having the $X$ shape when written in the basis of $\sigma_1\otimes\sigma_1$, which can be obtained from the standard one by applying the matrix $V\otimes V$, $V=\frac{\sigma_1+\sigma_3}{\sqrt{2}}$. The positivity condition are then readily obtained and read
\begin{align*}
	0\le\mu_1,\mu_2\le1 \, \qquad &0\le\nu\le 1-(\mu_1+\mu_2) \,, \qquad \abs{u}\le \sqrt{\mu_1 \mu_2} \,, \qquad \abs{v}\le \sqrt{\nu(1-\mu_1-\mu_2-\nu)}\,.
\end{align*}
Setting $\alpha_t=e^{-t}\cosh(t)$ and $\beta_t=e^{-2t}$, the states in~\eqref{mutualinfor2sequences} read
\begin{align*}
	&\begin{aligned}
		\Lambda_t\otimes\Lambda_t[\rho_X^{(1)}]&\,\,=\frac{1}{4}\big\{\mathds{1}_4-(1-2(\mu_1+\nu))\alpha_t\sigma_1\otimes\mathds{1}_2+(1-2(\mu_2+\nu))\alpha_t\mathds{1}_2\otimes\sigma_1  \\&\hskip0.5cm-(1-2(\mu_1+\mu_2))\alpha_t^2\sigma_1\otimes\sigma_1-2\Re(u-v)\alpha_t^2\sigma_2\otimes\sigma_2   \nonumber	\\&\hskip0.5cm+		2\Re(u+v)\beta_t^2\sigma_3\otimes\sigma_3+2\Im(u+v)\alpha_t\beta_t\sigma_2\otimes\sigma_3+2\Im(u-v)\alpha_t\beta_t\sigma_3\otimes\sigma_2\big\}\,,
	\end{aligned}\\
	&\begin{aligned}
		\Phi_{\bs{1}}\otimes\Phi_{\bs{1}}[\rho_X^{(1)}]&=\frac{1}{4}\big\{\mathds{1}_4-(1-2(\mu_1+\nu))\sigma_1\otimes\mathds{1}_2+(1-2(\mu_2+\nu))\mathds{1}_2\otimes\sigma_1\\&\hskip0.5cm-(1-2(\mu_1+\mu_2))\sigma_1\otimes\sigma_1-2\Re(u-v)\beta_t^2\sigma_2\otimes\sigma_2     \nonumber     \\&\hskip0.5cm+		2\Re(u+v)\beta_t^2\sigma_3\otimes\sigma_3+2\Im(u+v)\beta_t^2\sigma_2\otimes\sigma_3+2\Im(u-v)\beta_t^2\sigma_3\otimes\sigma_2\big\}\,,
	\end{aligned}\\
	&\begin{aligned}
		\Phi_{\bs{2}}\otimes\Phi_{\bs{2}}[\rho_X^{(1)}]&=\frac{1}{4}\big\{\mathds{1}_4-(1-2(\mu_1+\nu))\beta_t\sigma_1\otimes\mathds{1}_2+(1-2(\mu_2+\nu))\beta_t\mathds{1}_2\otimes\sigma_1\\&\hskip0.5cm-(1-2(\mu_1+\mu_2))\beta_t^2\sigma_1\otimes\sigma_1-2\Re(u-v)\sigma_2\otimes\sigma_2    \nonumber    \\&\hskip0.5cm+		2\Re(u+v)\beta_t^2\sigma_3\otimes\sigma_3+2\Im(u+v)\beta_t\sigma_2\otimes\sigma_3+2\Im(u-v)\beta_t\sigma_3\otimes\sigma_2\big\}\,,
	\end{aligned}\\
	&\begin{aligned}
		\Phi_{\bs{1}}\otimes\Phi_{\bs{2}}[\rho_X^{(1)}]&=\frac{1}{4}\big\{\mathds{1}_4-(1-2(\mu_1+\nu))\sigma_1\otimes\mathds{1}_2+(1-2(\mu_2+\nu))\beta_t\mathds{1}_2\otimes\sigma_1\\&\hskip0.5cm-(1-2(\mu_1+\mu_2))\beta_t^2\sigma_1\otimes\sigma_1-2\Re(u-v)\beta_t\sigma_2\otimes\sigma_2    \nonumber    \\&\hskip0.5cm+		2\Re(u+v)\beta_t^2\sigma_3\otimes\sigma_3+2\Im(u+v)\beta_t^2\sigma_2\otimes\sigma_3+2\Im(u-v)\beta_t\sigma_3\otimes\sigma_2\big\}\,,
	\end{aligned}\\
	&\begin{aligned}
		\Phi_{\bs{2}}\otimes\Phi_{\bs{1}}[\rho_X^{(1)}]&=\frac{1}{4}\big\{\mathds{1}_4-(1-2(\mu_1+\nu))\beta_t\sigma_1\otimes\mathds{1}_2+(1-2(\mu_2+\nu))\mathds{1}_2\otimes\sigma_1 \\&\hskip0.5cm-(1-2(\mu_1+\mu_2))\beta_t\sigma_1\otimes\sigma_1-2\Re(u-v)\beta_t\sigma_2\otimes\sigma_2\nonumber \\&\hskip0.5cm+		2\Re(u+v)\beta_t^2\sigma_3\otimes\sigma_3+2\Im(u+v)\beta_t\sigma_2\otimes\sigma_3+2\Im(u-v)\beta_t^2\sigma_3\otimes\sigma_2\big\}\,.
	\end{aligned}
\end{align*}

\section{Quantum Helstrom ensemble without entanglement}\label{app:noentanglementconstruction}
Consider the eternally non-Markovian  evolution $\Lambda_t$ generated by Pauli rates $\gamma_t(t)=1, \gamma_t^{(3)}=-\tanh(t)$. %RFor $s>0$, and $\epsilon>0$ very small, the application of $\Lambda_{s+\epsilon,s}\otimes \Lambda_{s+\epsilon,s}$ on the symmetric projector $P_+^2$ leads to
The symmetric projector $P_+^2$ always detects SBFI for Pauli second tensor powers and, for small $\epsilon>0$,
\begin{equation}
	\normt{\Lambda_{s+\epsilon,s}\otimes\Lambda_{s+\epsilon,s}[P_+^2]}\simeq 1+ 2 \epsilon\tanh(s)%-\frac{\epsilon^2}{2}\left(1+4\tanh(s)(1-\tanh(s))\right)
	>\normt{P_+^2}.
\end{equation}
for sufficiently small $\epsilon\ll s$. Now, we argue that there exists a Helstrom matrix 
\begin{equation}
	\Delta_\mu=\mu \rho_1-(1-\mu) \rho_2,
\end{equation}
with $\rho_1$, $\rho_2$ separable, such that $\Lambda_s\otimes \Lambda_s[\Delta_\mu]=\alpha P_{+}^2$ so that
\begin{equation}
	\normt{\Lambda_t\otimes\Lambda_t[\Delta_\mu]}=\alpha\normt{\Lambda_{t,s}\otimes\Lambda_{t,s}[P_+^2]}>\alpha \normt{P_+^2}= \normt{\Lambda_s\otimes \Lambda_s[\Delta_\mu]}.
\end{equation}
Consider the isotropic state
\begin{equation}
	\rho_a=(1-a) \frac{\mathds{1}_4}{4}+a P_2^+, \qquad 0\le a\le1,
\end{equation}
which is separable for $a\le1/3$.
The preimage of $P_+^2$ is
\begin{equation}
	\Lambda_s^{-1}\otimes \Lambda_s^{-1}[P_2^+]=\frac{1}{a}	\Lambda_s^{-1}\otimes \Lambda_s^{-1}[\rho_a]-\frac{1-a}{a} \frac{\mathds{1}_4}{4}
\end{equation}
Recall that $\Lambda_s^{-1}\otimes \Lambda_s^{-1}$ only guarantees hermiticity, but not, in general, positivity preservation. 
Nevertheless, for sufficiently small $a$, $\Lambda_s^{-1}\otimes \Lambda_s^{-1}[\rho_a]$ is separable by being sufficiently close to the separable state 
$\displaystyle \frac{\mathds{1}_4}{4}$.
Explicitly, in Fano form, 
\begin{equation}
	\rho_a=\frac{1}{4}\left[\mathds{1}_4+a(\sigma_1\otimes \sigma_1-\sigma_2\otimes\sigma_2+\sigma_3\otimes\sigma_3)\right].
\end{equation}
Given the Pauli eigenvalues  $\lambda_t=e^{-t}\cosh(t)$, $\lambda_t^{(3)}=e^{-2t}$, the algebraic inverse of $\rho_a$ is 
\begin{equation}
	\label{fanoinverse}
	\Lambda_a^{-1}\otimes\Lambda_a^{-1}[\rho_a]=\frac{1}{4}\left[\mathds{1}_4+a\frac{e^{2s}}{\cosh^2(s)}(\sigma_1\otimes \sigma_1-\sigma_2\otimes\sigma_2)+e^{4s}\sigma_3\otimes\sigma_3)\right].
\end{equation}
The matrix in \eqref{fanoinverse} is positive provided that $0\le a\le e^{-4s}$. 
Fix for instance $s=\textrm{arctanh}(\frac{1}{2})\approx0.55$. Then,
\begin{equation}
	\rho_a^0=\Lambda_{s}^{-1}\otimes\Lambda_{s}^{-1}[\rho_a]=\frac{1}{4}	\left(
	\begin{array}{cccc}
		1+9a & 0 & 0 & \frac{9 a}{2} \\
		0 &  1-9 a & 0 & 0 \\
		0 & 0 & 1-9 a & 0 \\
		\frac{9 a}{2} & 0 & 0 & 1+9 a \\
	\end{array}
	\right)
\end{equation}
is a physical state if $a<1/9$. Its partial transpose is
\begin{equation}
	T\otimes\mathrm{id}[\rho_a^0]=\frac{1}{4}	\left(
	\begin{array}{cccc}
		1+9a & 0 & 0 & 0 \\
		0 &  1-9 a & \frac{9 a}{2} & 0 \\
		0 & \frac{9 a}{2}& 1-9 a & 0 \\
		0 & 0 & 0 & 1+9 a \\
	\end{array}
	\right),
\end{equation}
which is positive for $a\le2/27\equiv a^*$. Hence, for $a\le a^*$, one has a well defined separable state $\rho_a^0$, such that 
\begin{equation}
	\normt{\Lambda_t\otimes \Lambda_t[\Delta_{\mu(a)}]}-\normt{\Lambda_{s}\otimes \Lambda_{s}[\Delta_{\mu(a)}]}>0
\end{equation}
for some $t> s$, where $\Delta_{\mu(a)}=\mu(a) \rho_a^0-(1-\mu(a))\frac{\mathds{1}_4}{4}$, with $\mu(a)=\frac{a}{1-a}$, $\rho_r^0$ separable. Since $\mathds{1}_4/4$ is a fully incoherent state with respect to every basis, the ensemble quantumness of correlations reduces to the geometric measure of quantum discord of the isotropic state,
\begin{equation*}
	\mathcal{Q}_{\{\mathbb{P}^1\otimes \mathbb{P}^2\}}({\chi^\mathcal{E}}(t))=\mu(a)\min_{\mathbb{P}^1\otimes\mathbb{P}^2} \normt{\rho_a-\mathbb{P}^1\otimes\mathbb{P}^2[\rho_a]}>0\,.
\end{equation*}

\end{appendices}

%\printbibliography
%\printbibitembibliography
\end{document}